\documentclass{article}
\usepackage{amsmath}
\usepackage{amssymb}
\usepackage{theorem}
\usepackage{xspace}
\usepackage[hyphens]{url}
\usepackage{datetime}
\usepackage{authblk}
\usepackage[usenames,dvipsnames,svgnames]{xcolor}

\usepackage{algorithmic, algorithm}
\usepackage{cite}
\usepackage[normal]{subfigure}
\usepackage{siunitx}
\usepackage{microtype}
\usepackage[margin=2.5cm]{geometry}

\usepackage{epstopdf}

\usepackage{multirow}
\usepackage{slashbox}

\usepackage{tikz}
\usetikzlibrary{shapes,arrows}
\usetikzlibrary{decorations,calc}

\newcommand{\prox}{\ensuremath{\operatorname{prox}}}

\newcommand{\FF}{\ensuremath{{\mathcal F}}}

\newcommand{\NN}{\ensuremath{\mathbb{N}}}
\newcommand{\RR}{\ensuremath{\mathbb{R}}}
\newcommand{\CC}{\ensuremath{\mathbb{C}}}

\newcommand{\RPP}{\ensuremath{\left]0,+\infty\right[}}

\newcommand{\bR}{\ensuremath{\mathbf{R}}}
\newcommand{\bh}{\ensuremath{\mathbf{h}}}

\newcommand{\bH}{\ensuremath{\mathbf{H}}}

\newcommand{\br}{\ensuremath{\mathbf{r}}}
\newcommand{\bb}{\ensuremath\mathbf{{b}}}
\newcommand{\bp}{\ensuremath\mathbf{{p}}}

\newcommand{\bq}{\ensuremath{\mathbf{q}}}
\newcommand{\bQ}{\ensuremath{\mathbf{Q}}}
\newcommand{\bB}{\ensuremath{\mathbf{B}}}

\newcommand{\bA}{\ensuremath{\mathbf{A}}}

\newcommand{\be}{\ensuremath{\mathbf{e}}}
\newcommand{\bE}{\ensuremath{\mathbf{E}}}
\newcommand{\bw}{\ensuremath{\mathbf{w}}}
\newcommand{\bP}{\ensuremath{\mathbf{P}}}
\newcommand{\bx}{\ensuremath{\mathbf{x}}}

\newcommand{\bz}{\ensuremath{\mathbf{z}}}

\newcommand{\hop}{\ensuremath{\mathsf{H}}}

\def\onevec{\mbox{1\hspace{-0.6em}1}}

\newcommand{\I}{\ensuremath{\operatorname{I}\xspace}}

\newcommand{\Diag}{\ensuremath{\operatorname{Diag}}}

\newtheorem{definition}{Definition}
\newtheorem{proposition}{Proposition}

\newcommand{\bv}{\vspace{-.01in}}

\newcommand{\bn}[1]{\textcolor{black}{#1}}
\newcommand{\new}[1]{\textcolor{black}{#1}}
\newcommand{\dom}{\ensuremath{\operatorname{dom}}}

\title{A Noise-Robust Method with Smoothed $\ell_1/\ell_2$ Regularization for Sparse Moving-Source Mapping}

\author[1]{Mai Quyen Pham \thanks{mai-quyen.pham@gipsa-lab.grenoble-inp.fr}}
\author[1,2]{Benoit Oudompheng \thanks{enoit.oudompheng@grenoble-inp.org}}
\author[1]{J\'er\^ome I. Mars \thanks{jerome.mars@gipsa-lab.grenoble-inp.fr}}
\author[1,3]{Barbara Nicolas \thanks{barbara.nicolas@creatis.insa-lyon.fr}}
\affil[1]{Universit\'e Grenoble-Alpes, GIPSA-lab, F-38000 Grenoble, France}
\affil[2]{MicrodB, 28 Chemin du Petit Bois, BP 36, 69131 \'Ecully Cedex, France }
\affil[3]{Universit\'e de Lyon, CREATIS, CNRS UMR5220; Inserm U1044; INSA-Lyon; Universit\'e Lyon 1, France}

\renewcommand\Authands{ and }

\begin{document}

\date{}
\maketitle

\begin{abstract}
The method described here performs blind deconvolution of the beamforming output in the frequency domain. To provide accurate blind deconvolution, sparsity priors are introduced with a smooth $\ell_1/\ell_2$ regularization term.
As the mean of the noise in the power spectrum domain is dependent on its variance in the time domain, the proposed method includes a variance estimation step, which allows more robust blind deconvolution.
Validation of the method on both simulated and real data, and of its performance, are compared with two well-known methods from the literature: the deconvolution approach for the mapping of acoustic sources, and sound density modeling.    
\end{abstract}
\begin{keywords}
Smoothed $\ell_1/\ell_2$ regularization, Sparse representation, Alternating minimization, Block coordinate, Proximal forward-backward,
Moving-source localization, Beamforming blind deconvolution, Acoustic signal processing, Robustness algorithms.
\end{keywords}

\section{Introduction}

\indent Blind deconvolution has a central role in the field of signal and image processing. \bn{It has many applications in communications \cite{Haykin_S_1994_book_blind_d}, nondestructive testing \cite{Nandi_A_1997_j-ieee-tsp_blind_dusnta}, image processing \cite{Kundur_D_1996_j-ieee-spm_blind_id,Kato_M_1999_j-ieice-tfeccs_set-theoretic_bidbhsdm,Ahmed_A_2014_j-ieee-tit_blind_ducp}, and medical imaging processing \cite{Campisi_P_2007_book_blind_idta}. Moreover, several acoustic issues can be formulated as blind deconvolution
 problems or blind source separation \cite{Zibulevsky_2001_j-neural-comput_blind_sssdsd}.} In many realistic scenarios, the blurring kernel (or the system) is imprecise or not known. Thus, the deconvolution problem becomes blind and underdetermined, and often requires additional hypotheses.

\bn{One possible additional hypothesis is the sparsity of the signal}, which is an extensively 
studied topic in signal processing. The main idea is to find the most compact representation 
of a signal that consists of only a few nonzero elements. \bn{In acoustic signal processing, sparsity can be introduced}, either in the system or the signal domain (input). 
In source localization \footnote{In this paper the terms ``source localization'' and ``source mapping'' are equivalent. They refers to the goal of the paper which is to map noise sources inside a global vehicle (here a boat) during a pass-by experiment.} in particular, \bn{the positions of sources can be considered as} sparsely distributed on a calculation grid. \bn{The question is then which measure can be used to evaluate the sparsity of a signal?} 
In \cite{Pereira_A_2013_thesis-acoustic_ies}, Pereira used $\ell_2^2$-norm as a penalty \bn{to stabilize inverse problem solutions}, which can be achieved using an adapted Tikhonov regularization method. 
However \bn{this penalty} is not adapted for the considered case of sparse source positions. An $\ell_1$-norm is popular to restore the sparsity of the solution, as proposed in \cite{Suzuki_T_2011_j-sound-vibration_L1generalized_ibfarcidms,Chu_L_2014_j-applied-acoustic_robust_srascai}.
However, in \cite{Benichoux_A_2013_p-icassp_fundamental_pbdssi}, Benichoux{\em~et al.} showed that use of the norm $\ell_1$ suffers from scaling and shift ambiguities due to the nonlinear relation between the blurring kernel and the signal, as also discussed in \cite{Comon_1996_j-ieee-spl_contrasts_mbd,Moreau_E_1997_j-ieee-spl_generlaized_cmbdls}.     
Felix{\em~et al.} extended this result for the case of $\ell_p,\,( p<1)$-norm in \cite{Esser_E_2015_p_resolving_sabdpf}. In particular, both of these reports showed that using the $\ell_1/\ell_2$ function can overcome this difficulty. 
In the present paper, we propose the use of the smoothed $\ell_1/\ell_2$ ratio mentioned in \cite{Reppetti_A_2014_j-ieee-spl_sparse_bdsl1l2r} to force the sparse representation of the signal in a blind deconvolution problem applied to moving-source mapping.    
 
This paper is organized as follows. Following this Introduction, Section~2 is devoted to a review of the related framework for \bn{moving-source mapping using deconvolution}. 
Section~3 presents the proposed forward model, and
Section~4 describes the minimization problem, the proposed algorithm, and some mathematical tools that are essential to this methodology. 
The performance of the proposed method is assessed in Section~5, where we detail the chosen optimization criteria and provide comparisons with two methods: the deconvolution approach for the mapping of acoustic sources (DAMAS-MS) and the sound density modeling (SDM) methods. The proposed methodology is first evaluated on realistic synthetic data, and then it is applied to \bn{real data recorded in Lake Castillon (Verdon Gorges, France)}. The conclusions and perspectives are drawn up in Section~6.

\section{Related work}

\indent In this section, we \bn{briefly present the classical} methods that have been developed \bn{for acoustic-source localisation}.  

\indent Many methods have been developed to solve this problem based on array processing. The most classical \bn{one} is beamforming \cite{Urick_R_1983_book_principles_us}, which has been extensively used due to its robustness against noise and environmental mismatch. 
However, classical beamforming cannot be used for pass-by experiments, \bn{where the `vehicle' is moving and the goal is to map the different acoustic noise sources in the vehicle}. 
Here instead, source mapping is achieved by the extension to beamforming for moving sources (BF-MS) \cite{Fleury_V_2011_j-asa_extension_dammas}.

Nevertheless, BF-MS spatial resolution is limited, as the image of a point source is the array transfer function, 
which is comprised of a main lobe and secondary lobes. \bn{Consequently, many improvements have been proposed to overcome this problem, including 
a hardware strategy to reduce the side-lobe levels, where the resolution of the main lobes is through optimization of the antenna geometry. 
In particular, several optimizations of the sensor positions of linear antennas were proposed through the use of pseudo-random distributions 
\cite{Vertatschitsch_E_1986_p-ieee_nonredundant_a,Moffet_A_1968_j-ieee-tap_minimum_rla,Smith_R_1970_j_constant_brabbss}. 
Furthermore, a numerical strategy classically uses the weighting coefficients, which shade the array aperture and thus taper the side lobes, 
and as a consequence, also enlarge the main lobe \cite{VeenVan_B_1988_j-ieee-asspm_beamforming_vasf}. 
Another common approach is to use deconvolution methods.}

Recently, Sijtsma proposed an extended version of the deconvolution method CLEAN \cite{Hogbom_J_1974_j-aas_aperture_snrdi,Tsao_J_1988_j-ieee-tsp_reduction_ssamict} \bn{for moving sources}, which is known as CLEAN-SC (\textit{i.e.}, CLEAN based on spatial-source coherence) \cite{Sijtsma_P_2007_j-ija_clean_bssc}. This has an approach similar to the matching pursuit method \cite{Mallat_S_1993_j-ieee-tsp_matching_ptfd}.
CLEAN-SC provides satisfactory results for high signal-to-noise ratios (SNRs), \bn{but} requires \textit{a-priori} knowledge of the number of sources \bn{which is not always known in practical cases}.    
Brooks and Humphreys developed another approach, known as DAMAS \cite{Brooks_T_2006_j-sound-vibration_deconvolution_amasdpma}, and its extensions \cite{Fleury_V_2011_j-asa_extension_dammas,Yardibi_T_2008_j-asa_sparsity_cdaasm}. These algorithms use the iterative Gauss-Seidel method for solving the linear inverse problem
under the nonnegative constraint on source powers.
A particular extension was dedicated to moving sources, as DAMAS-MS \cite{Fleury_V_2011_j-asa_extension_dammas}, which improves moving-source \bn{mapping} in a high SNR context.
Another popular method that was also developed for moving sources is the SDM method of \cite{Bruhl_S_2000_j-sound-vibration_acoustic_nsmbmam}, which is based on a gradient-descent optimization technique. 
This represented the first use of optimization techniques with a noise prior and constraints on the signal. 
\bn{These two methods (\textit{i.e.}, DAMAS-MS and SDM) will be used as the references for comparison with our proposed method.}
\newline \indent In the case of low SNRs, the problem is difficult to solve, \bn{and thus some other approaches need to be developed. \new{In array processing}, Swindlehurst and Kailath \cite{Swindlehurst_A_1992_j-ieee-tsp_performance_asbmpmema} proposed a 
first-order perturbation analysis of the multiple signal classification (MUSIC) and root-MUSIC algorithms for various model errors.} \bn{Another possibility is to include a} regularization term to stabilize the solution. 
The Tikhonov regularization is applied to jet noise-source localization \cite{Fleury_V_2008_p-aiaa_determination_admamu}. 
The sparse distribution of sources is also commonly used, as in \cite{Suzuki_T_2011_j-sound-vibration_L1generalized_ibfarcidms,Sun_K_2011_j-sensors_adaptive_srslgpe,Cotter_S_2005_j-ieee-tsp_sparse_slipmmv,Chu_L_2014_j-applied-acoustic_robust_srascai,
Malioutov_D_2005_j-ieee-tsp_sparse_rpslsa,Doisy_Y_2020_j-ieee-tsp_interference_ssabpsd,Chen_J_2005_p-icassp_adaptive_srslgpe}. These methods have been developed for fixed-source localization
and have not currently been extended to moving sources.

The goal of the present paper is to propose a new blind deconvolution method \bn{that is applied to} BF-MS results to improve moving-source \bn{mapping}. 
\newline \indent The strategy is to formulate the forward problem as an optimization problem, with constraints that are derived from the physical context. 
The proposed cost function contains several parts: (i) a data-fidelity term that accounts for the noise characteristics;
(ii) the smoothed $\ell_1/\ell_2$ ratio \cite{Reppetti_A_2014_j-ieee-spl_sparse_bdsl1l2r} that promotes \bn{sparsity in the} moving-source locations; and 
(iii) the knowledge of some physical properties \bn{of the} sources and \bn{the} system, and of the variance noise are also introduced, through the indicator functions. 

\section{Observation model}

\subsection{Beamforming for moving sources}
\begin{figure}[t!]
\centering
\includegraphics[width=8cm]{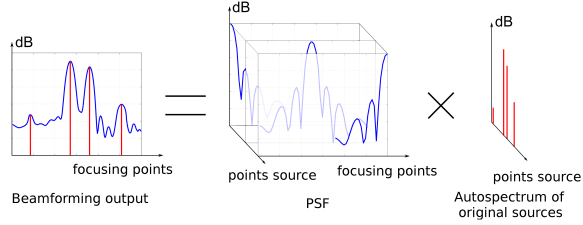}
\caption{Modeling of the forward problem\label{fig:deconv}}
\end{figure}
Beamforming for moving sources compensates for the Doppler effect and back-propagates the pressures measured by the $M$ sensor array to a calculation grid of $N$ points, which correspond to the possible source locations. 
We consider the classical case in pass-by experiments, in the far-field, \bn{with sources} that share the same \bn{global} movement and have low Mach numbers, 
$\|\overrightarrow{Ma}\| \ll 1$. 
For these conditions, some assumptions can be made over short time intervals of duration $T$, which are referred to as 
\emph{snapshots} \cite{Fleury_V_2011_j-asa_extension_dammas}, whereby:
\begin{itemize}
\item[1)] The sources are in fixed positions;
\item[2)] The Doppler effect is negligible at the frequencies and speeds of interest (\textit{i.e.}, it does not exceed the frequency resolution defined for the localization results).
\end{itemize}
Under these assumptions, BF-MS can be implemented in a simple way in the frequency domain. The measured acoustic pressures are temporally sliced into $K$ snapshots that are indexed by $k$. 
The calculation grid of $N$ points is defined for the snapshot $k$ \bn{using the \textit{a-priori} known global trajectory of the vehicle}. Note that this grid moves according to \bn{this trajectory}.

For the snapshot $k$, the pressures measured by these $M$ sensors at time $t\in\left[1,T\right]$ are denoted as $\check{\bp}_t^k \in\RR^M$, which can be divided into two parts:
\bn{\begin{equation}\label{eq:mesures}
\check{\bp}_t^k = \check{\overline{\bp}}_t^k + \check{\br}_t^k
\end{equation}}
in which $\check{\overline{\bp}}_t^k$ are the pressures measured by the $M$ sensors at time $t$ for the ideal case of free noise, and $\check{\br}_t^k$ is an additive noise in the recording domain. This defines the vectors:
\begin{align*}
\check{\bP}^k &= \left[(\check{\bp}_1^k)^\top,\,(\check{\bp}_2^k)^\top,\,\ldots,\, (\check{\bp}_T^k)^\top \right]^\top \in  \RR^{MT},\\
\check{\overline{\bP}}^k &= \left[(\check{\overline{\bp}}_1^k)^\top,\,(\check{\overline{\bp}}_2^k)^\top,\,\ldots,\, (\check{\overline{\bp}}_T^k)^\top \right]^\top \in  \RR^{MT},\\
\check{\bR}^k &= \left[(\check{\br}_1^k)^\top,\,(\check{\br}_2^k)^\top,\,\ldots,\, (\check{\br}_T^k)^\top \right]^\top \in  \RR^{MT},
\end{align*} 
where $\cdot^\top$ denotes the transpose.\\ 
We now consider the pressures measured in the frequency domain between $\zeta_1$ and $\zeta_F\,Hz$, which is related to a vector $\FF$ of length $F$. We define the Fourier transforms of $\check{\bP}^k,\, \check{\overline{\bP}}^k$ and $\check{\bR}^k$ for each snapshot $k$, as the following:
\begin{align*}
\bP^k & = \left[(\bp_1^k)^\top,\,(\bp_2^k)^\top,\,\ldots,\, (\bp_F^k)^\top \right]^\top \in  \CC^{MF}\\
\overline{\bP}^k &= \left[(\overline{\bp}_1^k)^\top,\,(\overline{\bp}_2^k)^\top,\,\ldots,\, (\overline{\bp}_F^k)^\top \right]^\top \in  \CC^{MF},\\
\bR^k &= \left[(\br_1^k)^\top,\,(\br_2^k)^\top,\,\ldots,\, (\br_F^k)^\top \right]^\top \in  \CC^{MF}
\end{align*}
where for every $f\in\{1,\,\ldots,\,F\}$, $\bp_f^k$, $\overline{\bp}_f^k$, and $\br_f^k$ are the vectors that contain the Fourier transform coefficients $p_f^k(m),\overline{p}_f^k(m)$, and $r_f^k(m)$ of the vectors 
$\left[\check{\bp}_1^k(m),\, \check{\bp}_2^k(m),\,\ldots,\,\check{\bp}_T^k(m)\right]^\top$, $\left[\check{\overline{\bp}}_1^k(m),\, \check{\overline{\bp}}_2^k(m),\,\ldots,\,\check{\overline{\bp}}_T^k(m)\right]^\top$, and 
$\left[\check{\br}_1^k(m),\, \check{\br}_2^k(m),\,\ldots,\,\check{\br}_T^k(m)\right]^\top$ at the frequency $\zeta_f \in \FF$ ($\zeta_f$ is the $f^{\text{th}}$ element of vector $\FF$), for every $m\in\{1,\,\ldots,\,M\}$, respectively.  

\bn{The BF-MS computed for the $n^{\text{th}}$ calculation point at the frequency $\zeta_f \in \FF$, and for the snapshot $k, b_{f}^k(n)$, is given by}: 
\begin{equation}\label{eq:bfms}
b_{f}^k(n) = |\left(\bw_{f,n}^k\right)^\hop \bp^k_{f}|^2
\end{equation}
where $\cdot^\hop$ is the conjugate transpose, and $\bw_{f,n}^k$ is the steering vector of length $M$ between the $M$ sensors and the $n^{\text{th}}$ calculation point. The $m^{\text{th}}$ element $w_{f,n}^k(m)$ of $\bw_{f,n}^k$ is:
\begin{equation}\label{eq:filtre}
w_{f,n}^k(m) = \left(\sum_{m'=1}^M \left(\frac{1}{d_{n,m'}^k}\right)^2\right)^{-1}\frac{\exp(-j\zeta_f d_{n,m}^k)}{d_{n,m}^k} 
\end{equation}
where $j$ is \bn{the square root of -1}, and $d_{n,m}^k$ is the distance between the $m^{\text{th}}$ sensor and the $n^{\text{th}}$ calculation point during the snapshot $k$. We then define the vector $\bb_{f} \in\RR^N$ with its $n^{\text{th}}$ element \bn{$b_{f}(n)$} as the estimate of the BF-MS output for the $n^{\text{th}}$ calculation point, through averaging over all of the $K$ snapshots; \textit{i.e.},
\begin{equation}
b_{f}(n) = \frac{1}{K}\sum_{k=1}^K b_{f}^k(n).
\end{equation}
Note that in the case considered, the \bn{receiver} array is a linear array along the $x$-axis. Consequently, BF-MS is performed along the $x$ dimension, and the calculation grid is a one-dimension vector of length $N$ along $x$. For more details on these computations, we refer the reader to \cite{Oudompheng_b_2015_p-oceans_passive_saainmms,Oudompheng_B_2015_thesis-localisation_csanptar}. 

\subsection{Inverse problem formulation}

We set the following assumptions:
\begin{itemize}
\item[(H1)]: The sources are random variables that are mutually independent and stationary;
\item[(H2)]: The number $M$ of the sensors is greater than the number $N_s$ of the sources (\textit{i.e.}, $M > N_s$), and these $N_s$ sources are sparsely distributed on the calculation grid;
\item[(H3)]: The noise components are mutually independent, and independent of the sources.
\end{itemize}
\bn{Using the expression of the BF-MS, and assuming that the sources are located on the $N$ points of the calculation grid, it is possible to express the BF-MS output at a given frequency $\zeta_f$, $\bb_f  \in \RR^N$, by}:
\begin{equation}
\bb_{f} = \bA_f  \overline{\bq}_{f}  + \bz_{f} 
\end{equation}
where $\bA_f \in\RR^{N\times N}$ (Fig.~\ref{fig:deconv}, middle) is the array transfer function matrix that contains the beamforming point-spread functions. The $(n,n') \in \{1,\,\ldots,\,N\}^2$ element $a_f(n,n')$ of $\bA_f$ is:
\begin{equation}\label{eqA}
a_f(n,n')	 = \frac{1}{K}\sum_{k=1}^K \left| \sum\limits_{m=1}^{M} \left(w_{f,n}^k(m)\right)^{\hop} \frac{\exp(-j\zeta_f d_{n',m}^k)}{d_{n',m}^k}  \right|^2
\end{equation}
$\bz_{f} \in \RR^{N}$ is the measurement noise,
 and $\overline{\bq}_{f} \in \RR^N$ is the autospectra of the possible sources located at the $N$ calculation points (Fig.~\ref{fig:deconv}, right), 
which \bn{are the unknowns to be estimated. This expression is frequently used in deconvolution, although it needs the knowledge of matrix $\bA_f$ related 
to the environment and to the array to perform the deconvolution. Nowadays, some research projects are focused on uncertain cases with partially known or unknown ocean environments \cite{Abadi_S_2012_j-acoust-sa_blind_drseasl}. 
For this reason, in this paper, we propose to formulate the BF-MS output at the frequency $\zeta_f$ as a blind deconvolution problem:}
\begin{equation}
\bb_{f} = \overline{\bh}_{f}\ast \overline{\bq}_{f} + \bz_{f}
\end{equation}
where $\overline{\bh}_{f}\in\RR^{P}$ is a blur unknown kernel, which needs to be estimated, as well as \bn{the autospectra of the sources}. $\ast$ denotes a discrete-time convolution operator (with appropriate boundary processing). 
\bn{Note that for deconvolution, the size $P$ of $\overline{\bh}_{f}$ has to be chosen by the user knowing the approximated size of the array transfer function.}\\
We now turn our attention to the term $\bz_{f}$, \bn{which} corresponds to the additional noise. 
In the literature, several methods have been \bn{proposed with $\bz_{f}$ as a Gaussian noise with zero mean (which is not adapted to the BF-MS signal). 
We propose to introduce the noise in the time recording domain and to model its transformation throught BF-MS. In acoustics, the noise components $\check{\br}^k_{m,t}$ in the time domain can commonly be} considered to be Gaussian, with zero mean and variance $\overline{\sigma}^2$. 
From Equations~\eqref{eq:mesures} and \eqref{eq:bfms}, and assumptions (H1) - (H3), we have: 
\begin{equation}\label{eq:8}
b_{f}(n) = \frac{1}{K} \sum_{k=1}^K \Big(|\left(\bw_{f,n}^k\right)^\hop\overline{\bp}^k_{f}|^2 + | \left(\bw_{f,n}^k\right)^\hop\br^{k}_{f}|^2\Big)
\end{equation}
\bn{Using Equation~\eqref{eq:8},} \bn{we assume in this paper that the observation noise} $\bz_{f}$ can be divided into two terms:
\begin{equation}
\bz_{f} =  \frac{1}{K}\left(\sum_{k=1}^K  \|\bw_{f,n}^k\|^2 \right)\overline{\sigma}^2 \onevec_{N} + \be_{f} 
\end{equation}
where $\|\cdot\|$ is the $\ell_2$-norm (which is also known as the Euclidean norm), \bn{$\onevec_{N}$ is a vector of ones of length $N$}, and \new{$\be_{f} \in \RR^{N}$ represents the remaining unknown effects,
the amplitude of these remaining effects is much lower than that of the variance $\overline{\sigma}^2$ of the Gaussian noise.} 
Note that:
\begin{equation*}
  \|\bw_{f,n}^k\|^2=  \left(\sum_{m=1}^M \left(\frac{1}{d_{n,m}^k}\right)^2\right)^{-1}
\end{equation*}
Consequently, the problem can be expressed with the following nonlinear problem in the standard form:
\begin{equation}\label{eq2}
\bB = \overline{\bH} \circledast \overline{\bQ} +  \overline{\sigma}^2 \delta \onevec_{NF}  +\bE 
\end{equation}
where 
\begin{align*}
\bB &= \left[\bb_1^\top,\,\bb_2^\top,\,\ldots,\,\bb_F^\top \right]^\top \in  \RR^{NF},\\
\overline{\bH} &= \left[\overline{\bh}_1^\top,\,\overline{\bh}_2^\top,\,\ldots,\,\overline{\bh}_F^\top \right]^\top \in  \RR^{PF},\\
\overline{\bQ} &= \left[\overline{\bq}_1^\top,\,\overline{\bq}_2^\top,\,\ldots,\,\overline{\bq}_F^\top \right]^\top \in  \RR^{NF},\\
\delta &= \frac{1}{K}\sum_{k=1}^K \left(\sum_{m=1}^M \left(\frac{1}{d^k_{n,m}}\right)^2\right)^{-1},\\
\bE &= \left[\be_1^\top,\,\be_2^\top,\,\ldots,\,\be_F^\top \right]^\top \in  \RR^{NF},
\end{align*}
and the discrete-time convolution operator $\circledast$ between $\overline{\bH}$ and $\overline{\bQ}$ is defined as follows:  
\begin{equation*}
\overline{\bH} \circledast \overline{\bQ} = \Big[(\overline{\bh}_{1} \ast \overline{\bq}_{1})^\top,\,(\overline{\bh}_{2} \ast \overline{\bq}_{2})^\top,\,\ldots,\,(\overline{\bh}_{F} \ast \overline{\bq}_{F})^\top\Big]^\top.
\end{equation*}

\section{Proposed method}
\subsection{Criterion to be minimized}
The purpose of this study is to identify $(\overline{\bH},\overline{\bQ},\overline{\sigma}^2)$ from $\bB$ through Equation~\eqref{eq2}, which leads to an inverse problem.  
To solve this, we propose an optimization \bn{approach} that minimizes the following criterion:  
\begin{equation}\label{Prob1}
\text{Find } (\widehat{\bH}, \widehat{\bQ}, \widehat{\sigma}^2) \in \underset{\bH \in \RR^{PF} , \bQ \in \RR^{NF}, \sigma^2 \in \RR_+}{\text{Argmin}}\; \theta(\bH,\bQ,\sigma^2)
\end{equation}
where:
\begin{equation}\label{eq_theta}
 \theta(\bH,\bQ,\sigma^2) = \phi(\bH,\bQ,\sigma^2)  + \varphi(\bQ) + \rho(\bH,\bQ,\sigma^2).
\end{equation}
The first term of Equation~\eqref{eq_theta}, $\phi : \RR^{PF} \times \RR^{NF} \times \RR_{+} \to \RR$ is a data fidelity term that is related to the observation model. In this case, we choose the least-squares objective function, \textit{i.e.},
\begin{equation}
\phi(\bH,\bQ,\sigma^2) = \frac{1}{2} \left\|\bH \circledast \bQ + \sigma^2 \delta \onevec_{NF} - \bB\right\|^2.
\end{equation}
The second term of Equation~\eqref{eq_theta}, $\varphi$, models a regularization function that accounts for the sparsity of the solution.  
In the present paper, we propose to use a new regularization function, the smoothed $\ell_1/\ell_2$ ratio, as proposed in \cite{Reppetti_A_2014_j-ieee-spl_sparse_bdsl1l2r};
\textit{i.e.}, for every $\bQ\in \RR^{NF}$, $(\lambda,\alpha,\beta,\eta) \in \RPP^4$: 

\begin{equation}
\varphi(\bQ) = \lambda \log\left( \frac{\ell_{1,\alpha}(\bQ) + \beta}{\ell_{2,\eta}(\bQ)} \right)
\end{equation}
with, 
\begin{align*}
\ell_{1,\alpha}(\bQ) &= \sum_{f=1}^F \sum_{n=1}^N \left( \sqrt{q_{f}(n)^2 + \alpha^2} - \alpha \right),	\\
\ell_{2,\eta}(\bQ) &= \sqrt{\sum_{f=1}^F \sum_{n=1}^N q_{f}(n)^2 + \eta^2}.
\end{align*}
The third term of Equation~\eqref{eq_theta}, $\rho: \RR^{PF} \times \RR^{NF} \times \RR_+ \to \RR$, is a regularization term that is related to some \textit{a-priori} constraints on the solution. In the following, we assume that $\rho$ can be split \bn{into three new terms that concern the three quantities to be estimated:}
\begin{equation*}
\rho(\bH,\bQ,\sigma^2) = \rho_1(\bH) + \rho_2(\bQ) + \rho_3(\sigma^2)
\end{equation*}
where $\rho_1$, $\rho_2$ and $\rho_3$ are (not necessarily smooth) proper, lower semicontinuous, convex functions, that are continuous on their domain, and which introduce the prior knowledge on the kernel blur (system), $\bH$, the source autospectra, $\bQ$, and the noise variance, $\sigma^2$. 
Due to these properties, the problem can be addressed with the block coordinate variable metric forward-backward algorithm \cite{Chouzenoux_E_2013_tr_block_cvmfba}. Moreover, in practice, $\bH$, $\bQ$ and $\sigma^2$ have different properties, and this choice \bn{allows} the \textit{a-priori} information to be taken into account independent of the \bn{searched quantities}.  

In the following, we denote: 
\begin{equation}
\psi(\bH,\bQ,\sigma^2) = \phi(\bH,\bQ,\sigma^2) + \varphi(\bQ).
\end{equation}

\subsection{Proposed algorithm}
The objective here is to provide a numerical solution to the optimization problem of Equation~\eqref{eq_theta}, which \bn{is a nonlinear blind deconvolution 
with three unknowns ($\bH, \bQ, \sigma^2$)}. One class of
popular solutions to solve this problem is the alternating minimization algorithm, 
by iteratively performing the three steps: (i) updating $\bH$ given $\bQ$ and $\sigma^2$; (ii) updating $\bQ$ given $\bH$ and $\sigma^2$; and (iii) updating $\sigma^2$ given $\bH$ and $\bQ$ \cite{Bolte_J_2010_p-icip_alternating_pabir}. Furthermore,
\bn{the criterion to minimize, which is formed as the sum of the smooth and nonsmooth functions, can be addressed with the block-variable metric by using an alternating forward-backward method \cite{Chouzenoux_E_2013_tr_block_cvmfba,Bolte_J_2013_j-math-prog-ser-a_proximal_almnnp}. 
This method combines explicitly the (forward) gradient step with respect to the smooth (not necessarily convex) functions and the proximal (backward) step with respect to the nonsmooth functions. 
The convergence of the algorithm can be accelerated using a majorize-minimize approach \cite{Chouzenoux_E_2014_j-optim_variable_mfbamsdfcf,Chouzenoux_E_2013_tr_block_cvmfba,Sotthivirat_S_2002_j-ieee-tip_image_rupspscaa}.}
 In this paper, we extend the smoothed one-over-two (SOOT) algorithm proposed in \cite{Reppetti_A_2014_j-ieee-spl_sparse_bdsl1l2r} by including a step for the noise variance estimation. 
\bn{This algorithm of noise-robust SOOT (NR-SOOT) is proposed, as presented in Algorithm~\ref{algo2}}. \bn{As previously mentioned, the block-variable metric forward-backward algorithm combines two steps of the process that requires two optimization principles. We now recall the definition of these}: The first is related to the choice of a variable metric that relies upon the majorization-minimization properties; \textit{i.e.},
\begin{definition}
Let $\psi: \RR^{N} \to \RR$ be a differentiable function. Let $x\in \RR^{N}$. Let us define, for every $x'\in \RR^{N}$:  
\begin{equation*}
\varrho(x',x) = \psi(x) + (x-x')^\top \nabla \psi(x) + \frac{1}{2} (x-x')^\top U(x)(x-x'), 
\end{equation*}
where $U(x) \in \RR^{N\times N}$ is a semidefinite positive matrix. Then, $U(x)$ satisfies the \bn{majoration} condition for $\psi$
at $x$ if $\varrho(\cdot,x)$ is a quadratic majorant of the function $\psi$ at $x$; \textit{i.e.}, for every $x'\in \RR^{N}$, $\psi(x') \leq \varrho(x',x)$.  
\end{definition}
A function $\psi$ has a $\mu$-Lipschitzian gradient on a convex subset $C\in\RR^{N}$, with $\mu>0$, if for every $(x,x')\in C^{2},\, \| \nabla \psi (x) -  \nabla \psi (x') \| \leq \mu \|x-x'\|$. Then, for every $\bx\in C$, a quadratic majorant of $\psi$ at $\bx$ is easily obtained taking $U(\bx) = \mu \I_N$, \bn{where $\I_N$ is the identity matrix of $\RR^{N\times N}$}.

The second optimization principle is the definition of the proximity operator of a proper, lower semicontinuous, convex function, relative to the metric induced by a symmetric positive definite matrix, which is defined in \cite{Combettes_P_2013_j-non-linear-anal_variable_mqfm} as follows:
\begin{definition}
Let $\rho:\RR^{N} \to ]-\infty,+\infty]$ be a proper, lower semicontinuous, and convex
function, let $U \in \RR^{N\times N}$ be a symmetric positive definite matrix, and let $x \in \RR^N$. The proximity operator of $\rho$ at $x$ relative to the metric induced by $U$ is the unique minimizer of $\rho + \frac{1}{2}(\cdot - x)^\top U (\cdot - x)$, and it is denoted by $\prox_{U,\rho}(x)$. If $U$ is equal to $\I_N$, then $\prox_{\rho} := \prox_{\I_N,\rho}$ is the proximity operator originally defined in \cite{Combettes_P_2011_book-proximal_smsp}.
\end{definition}

\begin{algorithm*}
\caption{The NR-SOOT algorithm. \label{algo2}}
\begin{algorithmic}
\STATE  For every $l \in \NN$ , let $I_l \in \NN^*$, $J_l \in \NN^*$. Let $(\gamma_1^{l,i})_{0 \le i \le I_l-1}$, $ (\gamma_2^{l,j})_{0 \le j \le J_l-1}$, and $\gamma_3^{l}$ be positive sequences. 
Initialize with  $\bH^0 \in \dom(\rho_1)$, $\bQ^0 \in \dom(\rho_2)$, and $\sigma^{2,0} \in \dom(\rho_3)$.
\STATE \textbf{Iterations:} 
\STATE
$	
\begin{array}{l}
\text{For } l = 0,1,\ldots	\\
\left\lfloor               
\begin{array}{l} 
	\bQ^{l,0} = \bQ^l , \, \,  \bH^{l,0} =  \bH^l , \\	
	\text{For } i = 0, \ldots, I_l-1\\
	\left\lfloor
	\begin{array}{l} 
		\widetilde{\bH}^{l,i+1} =  \bH^{l,i} - \gamma_1^{l,i}G_1(\bH^{l,i},\bQ^{l},\sigma^{2,l})^{-1} \nabla_1\psi (\bH^{l,i},\bQ^{l},\sigma^{2,l}) \\
		\bH ^{l,i+1} = \prox_{(\gamma_1^{l,i})^{-1}G_1(\bH^{l,i},\bQ^{l},\sigma^{2,l}),\rho_1} (\widetilde{\bH}^{l,i+1})
	\end{array} 
	\right.   \\
	\bH^{l+1} = \bH^{l,I_l}\\	
	\text{For } j = 0, \ldots, J_l-1\\
	\left\lfloor
	\begin{array}{l} 
	\widetilde{\bQ}^{l,j+1} =  \bQ^{l,j} - \gamma_2^{l,j} G_2(\bH^{l+1},\bQ^{l,j},\sigma^{2,l})^{-1} \nabla_2\psi (\bH^{l+1}, \bQ^{l,j},\sigma^{2,l}) \\
	\bQ^{l,j+1}= \prox_{(\gamma_2^{l,j})^{-1}G_2(\bH^{l+1},\bQ^{l,j},\sigma^{2,l}),\rho_2}(\widetilde{\bQ}^{l,j+1}) 
	\end{array}
	\right.   \\
	\bQ^{l+1} = \bQ^{l,J_l}	\\
 \widetilde{\sigma}^{2,l} =  \sigma^{2,l} - \gamma_3^l G_3(\bH^{l+1},\bQ^{l+1},\sigma^{2,l})^{-1} \nabla_3\psi (\bH^{l+1},\bQ^{l+1},\sigma^{2,l}) \\
	\sigma^{2,l+1} = \prox_{(\gamma_3^l)^{-1}G_3(\bH^{l+1},\bQ^{l+1},\sigma^{2,l}),\rho_3} \left( \widetilde{\sigma}^{2,l} \right)  
\end{array}     
\right.   \\	   
\end{array}
$
\end{algorithmic}
\end{algorithm*}

In this algorithm, \bn{$\nabla_1, \nabla_2$, and $\nabla_3$ are the partial gradients of $\psi$ with respect to the variables $\bH,\bQ$, and $\sigma^2$}. $G_1,\,G_2$, and $G_3$ are the semidefinite positive matrix used \bn{to build} the majorizing approximations of $\psi$ with respect to $\bH,\,\bQ$, and $\sigma^2$, and their expressions are given by the following proposition, as established in \cite{Reppetti_A_2014_j-ieee-spl_sparse_bdsl1l2r}:
\begin{proposition}
For every $(\bH,\bQ,\sigma^2) \in \RR^{PF} \times \RR^{NF} \times \RR_{+}$, let:
\begin{align*}
&G_1(\bQ,\bH,\sigma^2) = \mu_1(\bQ,\sigma^2)  \operatorname{I}_{PF} ,\\
&G_2(\bQ,\bH,\sigma^2) = \left( \mu_2(\bH,\sigma^2) + \frac{9\lambda}{8\eta^2} \right) \operatorname{I}_{NF}\\ 
&\hspace{4.5cm} + \frac{\lambda}{\ell_{1,\alpha}(\bQ)+\beta} G_{\ell_{1,\alpha}}(\bQ),\\
& G_3(\bQ,\bH,\sigma^2) = \mu_3(\bH,\bQ) ,
\end{align*}
where:
\begin{equation}	\label{eq:maj_l1s}
	G_{\ell_{1,\alpha}}(\bQ) = \Diag \left( \left( (q_{f}(n)^2 + \alpha^2)^{-1/2} \right)_{1 \le f \le F,\,1 \le n \le N} \right) ,
\end{equation} 
and $\mu_1(\bQ,\sigma^2)$, $\mu_2(\bH,\sigma^2)$, and $\mu_3(\bH,\bQ)$ are the Lipschitz constants for
$\nabla_1\phi( \cdot, \bQ,\sigma^2)$, $\nabla_2 \phi(\bH, \cdot, \sigma^2)$, and $\nabla_3 \phi(\bH,\bQ, \cdot)$, respectively.\footnote{Such Lipschitz constants are straightforward to derive since
$\phi$ is a quadratic cost.} 
Then, $G_1(\bH,\bQ,\sigma^2)$,  $G_2(\bH,\bQ,\sigma^2))$, and $G_3(\bH,\bQ,\sigma^2)$ satisfy the majoration condition for $\psi(\cdot,\bQ,\sigma^2)$ at $\bH$, $\psi(\bH,\cdot,\sigma^2)$ at $\bQ$, and $\psi(\bH,\bQ,\cdot)$ at $\sigma^2$, respectively.
\end{proposition}
To conclude, we have proposed a blind deconvolution method to apply to the BF-MS that imposes sparsity \bn{on the noise acoustic-source locations}. This method is validated in the next section, and compared to the \bn{classical methods} of DAMAS-MS and SDM, used in acoustics for moving-source deconvolution.

\section{Results}
\begin{figure}[h!]
\centering
\begin{tikzpicture}[scale=1.2,>=latex,x=1.2cm,y=1cm]
	\draw[->,line width=0.1mm] (-3,0) -- (3,0) node[below left]{\footnotesize $x$};
	\draw[->,line width=0.1mm] (0,-1.2) -- (0,1) node[below left]{\footnotesize $z$};
	\draw[->,line width=0.1mm] (-0.8,-0.8) -- (1,1) node[below]{\footnotesize $y$};
	\draw [fill=red] (-2.6,-0.1) rectangle (-1,0.1);
	\draw[-,line width=0.1mm] (-2.4,-0.1) -- (-2.4,0.1);
	\draw[-,line width=0.1mm] (-2.2,-0.1) -- (-2.2,0.1);
	\draw[-,line width=0.1mm] (-2.0,-0.1) -- (-2.0,0.1);
	\draw[-,line width=0.1mm] (-1.8,-0.1) -- (-1.8,0.1);
	\draw[-,line width=0.1mm] (-1.6,-0.1) -- (-1.6,0.1);
	\draw[-,line width=0.1mm] (-1.4,-0.1) -- (-1.4,0.1);
	\draw[-,line width=0.1mm] (-1.2,-0.1) -- (-1.2,0.1);
	\draw[-,line width=0.1mm] (-1.0,-0.1) -- (-1.0,0.1);
	\draw [fill=black] (-2.4,-0.1) rectangle (-2.2,0.1);
	\draw [fill=green] (-2.0,-0.1) rectangle (-1.8,0.1);
	\draw[dotted] (-1,-0.1) -- (-1,-0.9);
	\draw[dotted] (-2.6,-0.1) -- (-2.6,-0.9);

	\draw(-2.7,0.3) node {\footnotesize $X_1(0)$};
	\draw(-1.1,0.3) node {\footnotesize $X_N(0)$};
	\draw(-2.3,-0.3) node {\footnotesize $S_1(0)$};
	\draw(-1.7,-0.3) node {\footnotesize $S_2(0)$};
	\draw[<->,line width=0.1mm] (-2.4,0.4) -- (-1.8,0.4);
	\draw(-2.1, 0.6) node {\footnotesize $5m$};
	\draw[<->,line width=0.1mm] (-2.6,-0.9) -- (-1,-0.9);
	\draw(-1.8, -1.1) node {\footnotesize $20m$};
	\draw[blue,->,line width=0.5mm] (-0.5,1.1) -- (0.5,1.1);
	\draw[blue](-0.1, 1.3) node {\Large $v$};
	
	\draw [fill=red] (1,-0.1) rectangle (2.6,0.1);
	\draw[-,line width=0.1mm] (2.4,-0.1) -- (2.4,0.1);
	\draw[-,line width=0.1mm] (2.2,-0.1) -- (2.2,0.1);
	\draw[-,line width=0.1mm] (2.0,-0.1) -- (2.0,0.1);
	\draw[-,line width=0.1mm] (1.8,-0.1) -- (1.8,0.1);
	\draw[-,line width=0.1mm] (1.6,-0.1) -- (1.6,0.1);
	\draw[-,line width=0.1mm] (1.4,-0.1) -- (1.4,0.1);
	\draw[-,line width=0.1mm] (1.2,-0.1) -- (1.2,0.1);
	\draw[-,line width=0.1mm] (1.0,-0.1) -- (1.0,0.1);
	\draw [fill=black] (1.2,-0.1) rectangle (1.4,0.1);
	\draw [fill=green] (1.6,-0.1) rectangle (1.8,0.1);
	
	\draw(0.9,0.3) node {\footnotesize $X_1(D)$};
	\draw(2.5,0.3) node {\footnotesize $X_N(D)$};
	\draw(1.2,-0.3) node {\footnotesize $S_1(D)$};
	\draw(1.9,-0.3) node {\footnotesize $S_2(D)$};
	
	\draw (-0.5,-1.5) node {$\bullet$} (-0.3,-1.5) node {$\bullet$} (-0.1,-1.5) node {$\bullet$} (0.2,-1.5) node {$\cdots$} (0.5,-1.5) node {$\bullet$};
	
	\draw(0, -1.8) node {21 hydrophones};
	\draw[<->,line width=0.1mm] (-0.6,-2.1) -- (0.6,-2.1);
	\draw(0, -2.3) node {\footnotesize $10m$};
	\draw[<->,line width=0.1mm] (-0.3,-0.4) -- (-0.3,-1.3);
	\draw(-0, -0.8) node {\footnotesize $10m$};
\end{tikzpicture}
\caption{Simulated configuration of a pass-by experiment. Black, source $S_1$; green, source $S_2$; red, calculation grid; blue arrow, global movement of the sources.
\label{fig:schema}}
\end{figure}
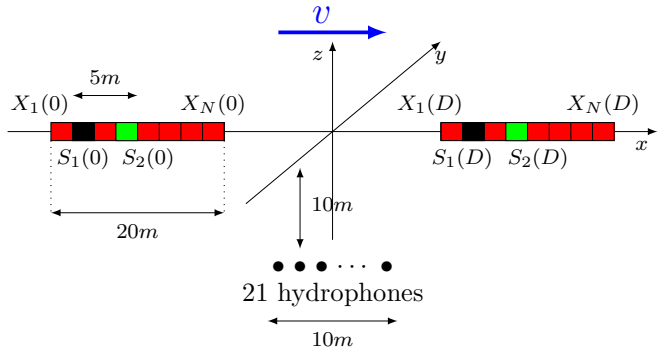

We consider synthetic and real data for the method validation. The synthetic data \bn{allow the use of quantitative indicators, whereas real data only provide subjective results.} For both cases, we perform comparative
evaluation with the standard algorithms of DAMAS-MS and SDM. In practice, the kernel blur related to the array transfer function has finite energy, and thus $\rho_1$ can be chosen as an indicator function of set $C = \left\{\bH \in [h_{\min},h_{\max}]^{PF}\; | \; \|\bH\| \leq \kappa\right\}$ (equal to $0$ if $\bH\in C$, and $+\infty$ otherwise), where $\kappa >0$, and $h_{\min}$ and $h_{\max}$ are the minimum and maximum values of $\overline{\bH}$, respectively. In the real data case, we choose $h_{\min} = 0$ and $h_{\max} = 1$. As mentioned before, the autospectra of sources $\bQ$ is sparse; moreover, it is limited in amplitude. Then, one natural choice for $\rho_2$ is the indicator function of the hypercube $[q_{\min},q_{\max}]^{NF}$,
where $q_{\min}$ (resp. $q_{\max}$) is the lower (resp., upper) boundary of $\overline{\bQ}$. In practice, we choose $q_{\min}$ as $0$, which leads to a nonnegative constraint on the source power variables,  and $q_{\max}$ is the maximum value of $\bB$.
Finally, the function $\rho_3$ related to the constraint on the noise variance, is equal to the indicator function of the interval $[\sigma^2_{\min},\sigma^2_{\max}]$, where $\sigma^2_{\min} = 0$ and $\sigma^2_{\max} =1$.        

The NR-SOOT algorithm with the penalty smoothed $\ell_1/\ell_2$ function and the classical DAMAS-MS and SDM algorithms are applied to the BF-MS result. For every $l\in \NN$, the number of inner-loops are fixed as $I_l = 1$ and $J_l = 100$.
The NR-SOOT algorithm is launched on $5000$ iterations, and it can stop earlier at iteration $l$ if $\|\bQ^{l} - \bQ^{l-1}\| \leq \sqrt{NF} \times 10^{-6}$. 

\subsection{Synthetic data}
The simulated configuration is presented in Figure~\ref{fig:schema}. Here, we consider two sources: 
a random broadband source located at $S_1 = (-4\,m,\,0\,m,\,0\,m)$ (Fig.~\ref{fig:schema}, black) and a sum of $3$ sine functions at frequencies $1200\, Hz$, $1400\, Hz$, and $1800\, Hz$ located at $S_2 = (1\,m,\,0\,m,\,0\,m)$ (Fig.~\ref{fig:schema}, green), \new{in the coordinate system whose origin is the center of the moving calculation grid all the time}.
The sources are moving jointly, and follow a linear trajectory of length $20\, m$ at constant speed $v = 2\,m/s$. A linear antenna of $21$ hydrophones that are equally spaced \bn{(with an inter-sensor distance of $0.5\,m$)} records the propagated acoustic signals over $D=10\, s$.
Zero-mean white Gaussian noise is added to the recorded signals. 
To perform BF-MS, the moving calculation grid $X_n(t),\, \forall n\in \{1,\,\ldots,\,N\}$ has a length of $20\,m$ and contains $N = 101$ points.

\begin{figure}[h!]
\centering
\includegraphics[width=8cm, height=6cm]{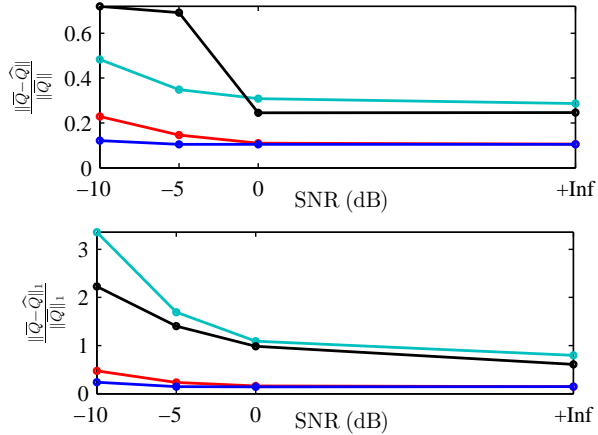}
\caption{Comparison of the results obtained by the DAMAS-MS (greenish-blue), SDM (black), SOOT original (red), and NR-SOOT (blue) algorithms, for input data of free noise and three different noise levels of the SNR $\in\{-10,\,-5,\,0\}$ dB.\label{tab}}
\end{figure}


Figure~\ref{tab} summarizes the quantitative results in terms of the reconstruction error $\overline{\bQ}$ for an input without noise and then with three different noise levels of the SNR, $\in\{-10,\,-5,\,0\}$ dB. The relative error is defined as the $\|\cdot\|$-norm (Fig.~\ref{tab}, top) and $\|\cdot\|_1$-norm (Fig.~\ref{tab}, bottom) between the real $\overline{\bQ}$ and the estimated $\widehat{\bQ}$, which demonstrates that the method can reconstruct accurately in terms of the amplitude energy and the sparse source positions (\textit{i.e.}, the smaller, the better). 
From Figure~\ref{tab}, we observe that SDM \bn{performs better} than DAMAS-MS \bn{for the case considered} (the $\|\cdot\|_1$-norm values of the residual error by SDM are always smaller than those by DAMAS-MS). However, \bn{the performance of SDM decreases significantly when the SNR decreases}. The SOOT original \bn{provides very satisfying results compared} to DAMAS-MS and SDM. Its \bn{performances} for cases of high SNR are similar to the proposed method, but for the cases of low SNR, the NR-SOOT algorithm \bn{is the only one that provides a satisfactory source-location estimation}.
Consequently, in all of these cases, the NR-SOOT algorithm has the smallest error \bn{for the source-location estimation}.

\begin{figure}[h!]
\centering
\subfigure[Input data without noise]{\includegraphics[width=7cm]{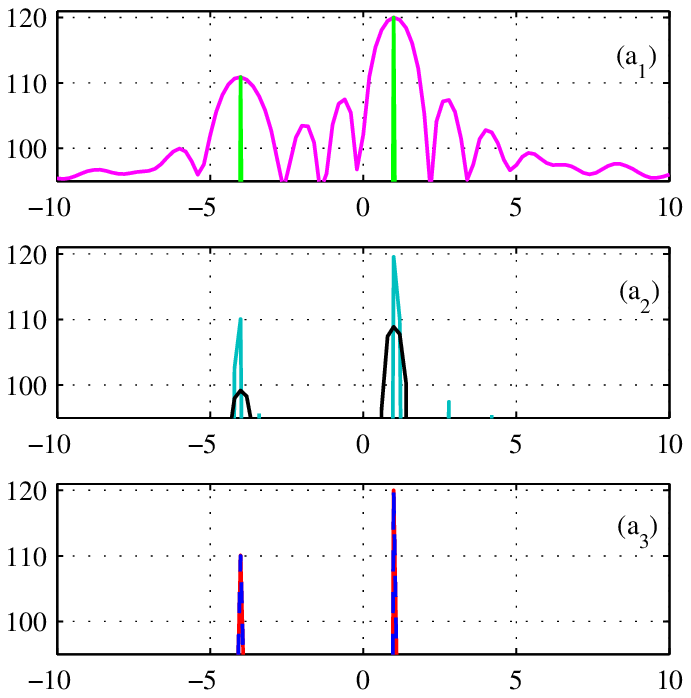}}
\subfigure[Input data with SNR of -5 dB]{\includegraphics[width=7cm]{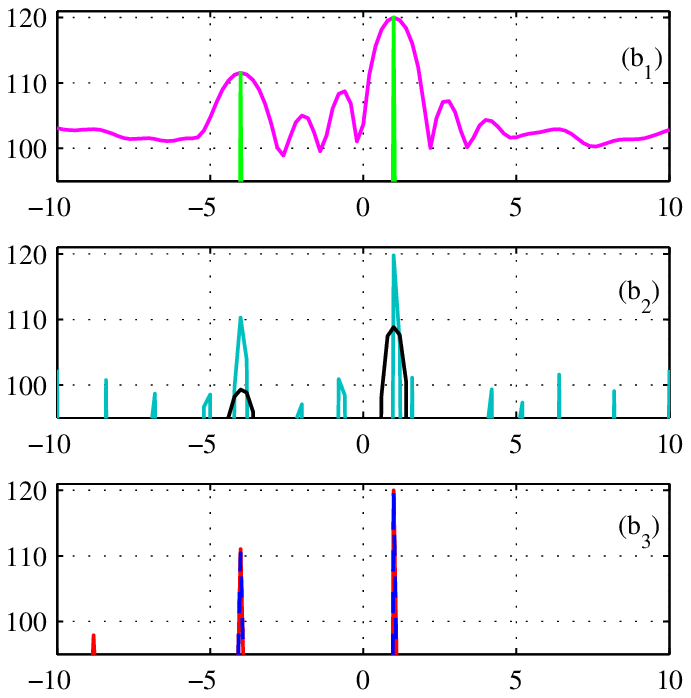}}
\caption{Top: Autospectrum of the original sources (green), with the BF-MS output (magenta). Middle: Results obtained using DAMAS-MS (greenish-blue) \cite{Fleury_V_2011_j-asa_extension_dammas}, and SDM (black) \cite{Bruhl_S_2000_j-sound-vibration_acoustic_nsmbmam}. Bottom: Results obtained using SOOT original (red), 
and NR-SOOT (dashed thick blue), at the frequency of $1400\, Hz$, without noise (left), and with SNR of -5 dB (right). \label{random_cosinus_-14_-9_1D_1400}}
\end{figure}

\begin{figure}[h!]
\centering
\subfigure[Input data without noise]{\includegraphics[width=7cm]{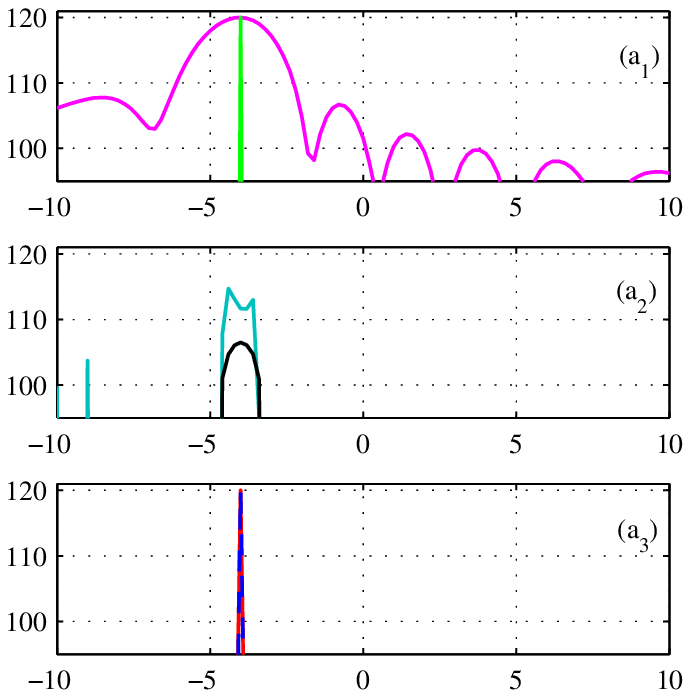}}
\subfigure[Input data with SNR of -5 dB]{\includegraphics[width=7cm]{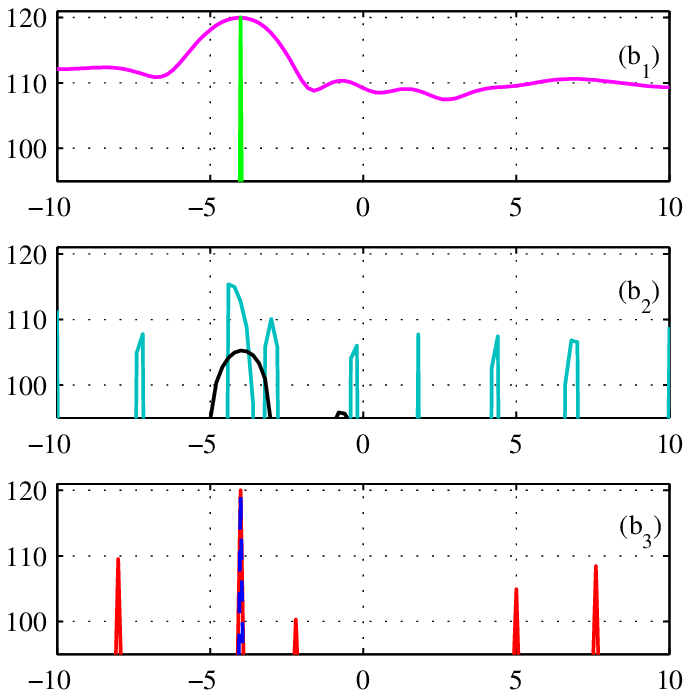}}
\caption{Top: Autospectrum of the original sources (green), with the BF-MS output (magenta). Middle: Results obtained using DAMAS-MS (greenish-blue) \cite{Fleury_V_2011_j-asa_extension_dammas}, and SDM (black) \cite{Bruhl_S_2000_j-sound-vibration_acoustic_nsmbmam}. Bottom: Results obtained using SOOT original (red), and NR-SOOT (dashed thick blue), at the frequency of $770\, Hz$, without noise (left), and with SNR of  -5 dB (right). \label{random_cosinus_-14_-9_1D_770}}
\end{figure}

After this quantitative study, it is necessary to investigate the performances of these methods qualitatively, directly on the localization maps for input data without noise and for a SNR of -5 dB. 
Figure~\ref{random_cosinus_-14_-9_1D_1400} and Figure~\ref{random_cosinus_-14_-9_1D_770} show the results for the DAMAS-MS, SDM, SOOT original, and NR-SOOT algorithms at frequencies of $1400\,Hz$ and $770\,Hz$, 
respectively. In these Figures, the green lines (a$_1$,b$_1$) represent the \bn{theoretical sources to estimate} (in terms of position and amplitude), 
the magenta lines represent the BF-MS results, which are the starting points of the DAMAS-MS, SDM, SOOT original, and SR-SOOT algorithms. 
In Figure~\ref{random_cosinus_-14_-9_1D_1400} and Figure~\ref{random_cosinus_-14_-9_1D_770}, the results obtained by DAMAS-MS are in greenish-blue, those of SDM are in black (a$_2$,b$_2$), those of SOOT original are in red, and 
those of NR-SOOT are in blue (a$_3$,b$_3$).

At the frequency of $1400\,Hz$ (Fig.~\ref{random_cosinus_-14_-9_1D_1400}), for which both sources exist, for the case without noise \bn{on the recorded data} (Fig.~\ref{random_cosinus_-14_-9_1D_1400}a)
both the SOOT original and the NR-SOOT algorithms detect the source positions accurately. 
DAMAS-MS gives some false alarms at $x = -3\, m$ and $x = 2.5\,m$. These false sources have small amplitudes, but they are a real problem because 
the number of sources is generally unknown. \bn{SDM locates two sources, but the amplitude estimation is not satisfactory and these sources are spread
in the space}. For the case of a SNR of -5 dB, DAMAS-MS \bn{does not succeed at all}, and it shows several false alarms with significant amplitudes. {Similar to the case without noise, with the SDM method, although 
there is no false alarm, again, the amplitudes are not correct}. The SOOT original algorithm gives good results, although there is one false alarm around $x = -9\,m$. In contrast, the NR-SOOT algorithm gives perfect results in terms of localization and source amplitude estimation.    

We now turn our attention to the case at the low frequency of $770\,Hz$ (Fig.~\ref{random_cosinus_-14_-9_1D_770}), for which only the source $S_1$ exists. In this case, DAMAS-MS gives a wrong result, with a spatially extended source and false alarms \bn{even in the noise-free case}. 
SOOT original \bn{gives satisfactory results for the free noise (Fig.~\ref{random_cosinus_-14_-9_1D_770}a$_3$), although when the SNR decreases}, the SOOT original algorithm creates false alarms (Fig.~\ref{random_cosinus_-14_-9_1D_770}b$_3$),
while the NR-SOOT algorithm shows excellent results in terms of position and amplitude (Fig.~\ref{random_cosinus_-14_-9_1D_770}b$_3$).
 \bn{The NR-SOOT algorithm is robust against noise.} In the following, for the sake of simplicity, and as it always provides the best results, we only consider the \bn{NR-SOOT algorithm for blind deconvolution of the two-dimensional illustrations}.

\begin{figure}[h!]
\centering
\subfigure[Initial BF-MS ]{\includegraphics[width=7.5cm, height=5cm]{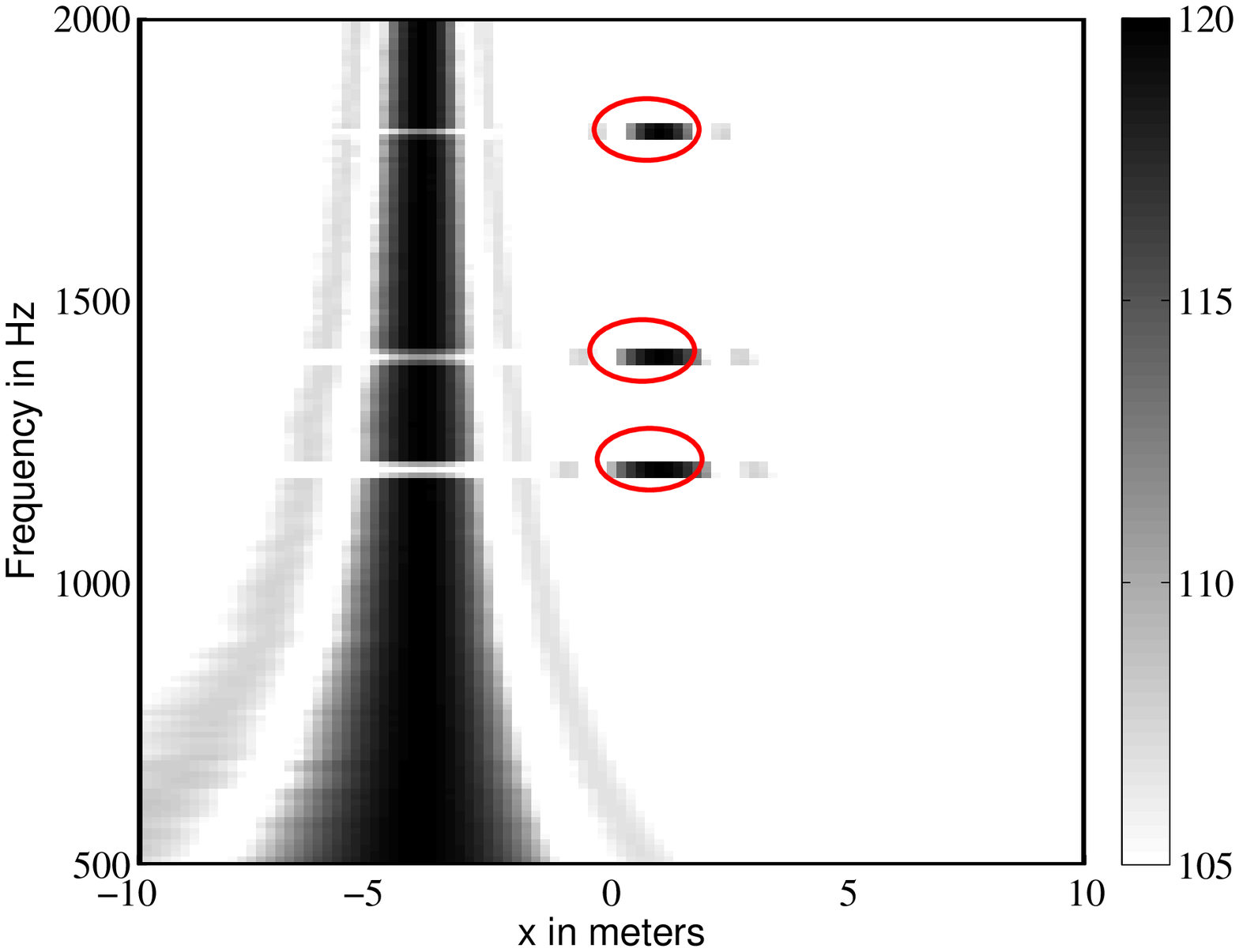}}
\subfigure[Deconvolution with DAMAS-MS]{\includegraphics[width=7.5cm, height=5cm]{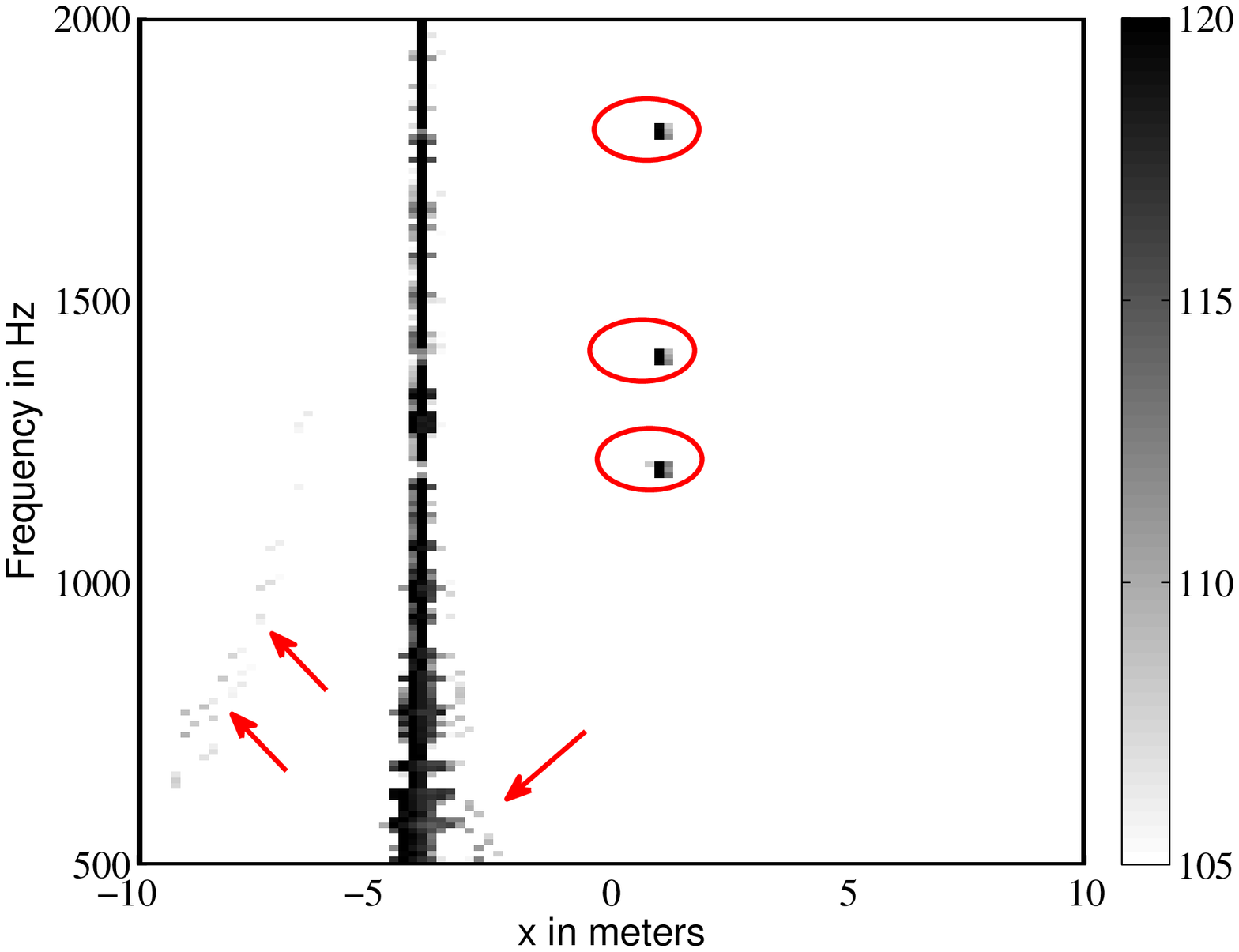}}
\subfigure[Deconvolution with SDM]{\includegraphics[width=7.5cm, height=5cm]{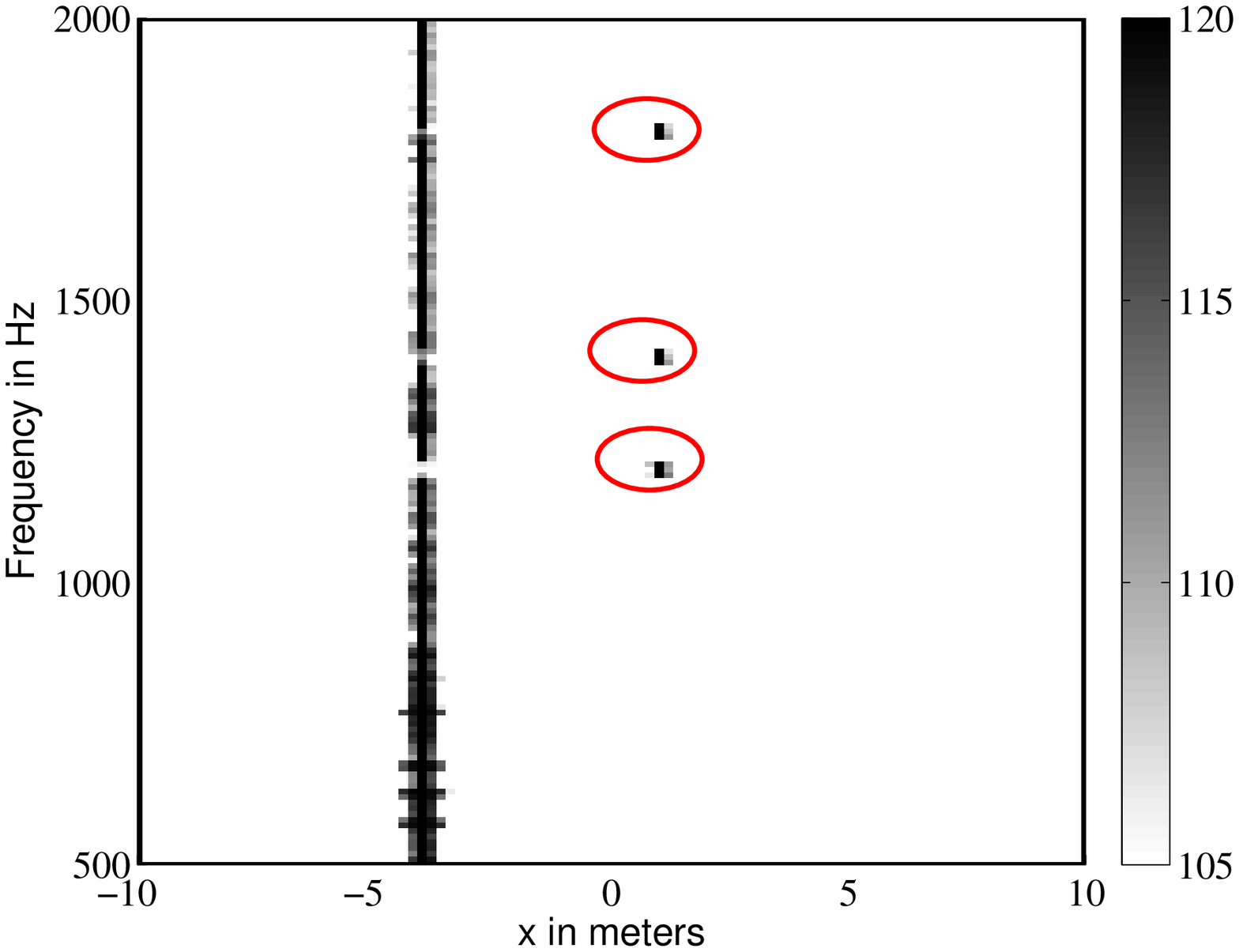}}
\subfigure[Blind deconvolution with NR-SOOT]{\includegraphics[width=7.5cm, height=5cm]{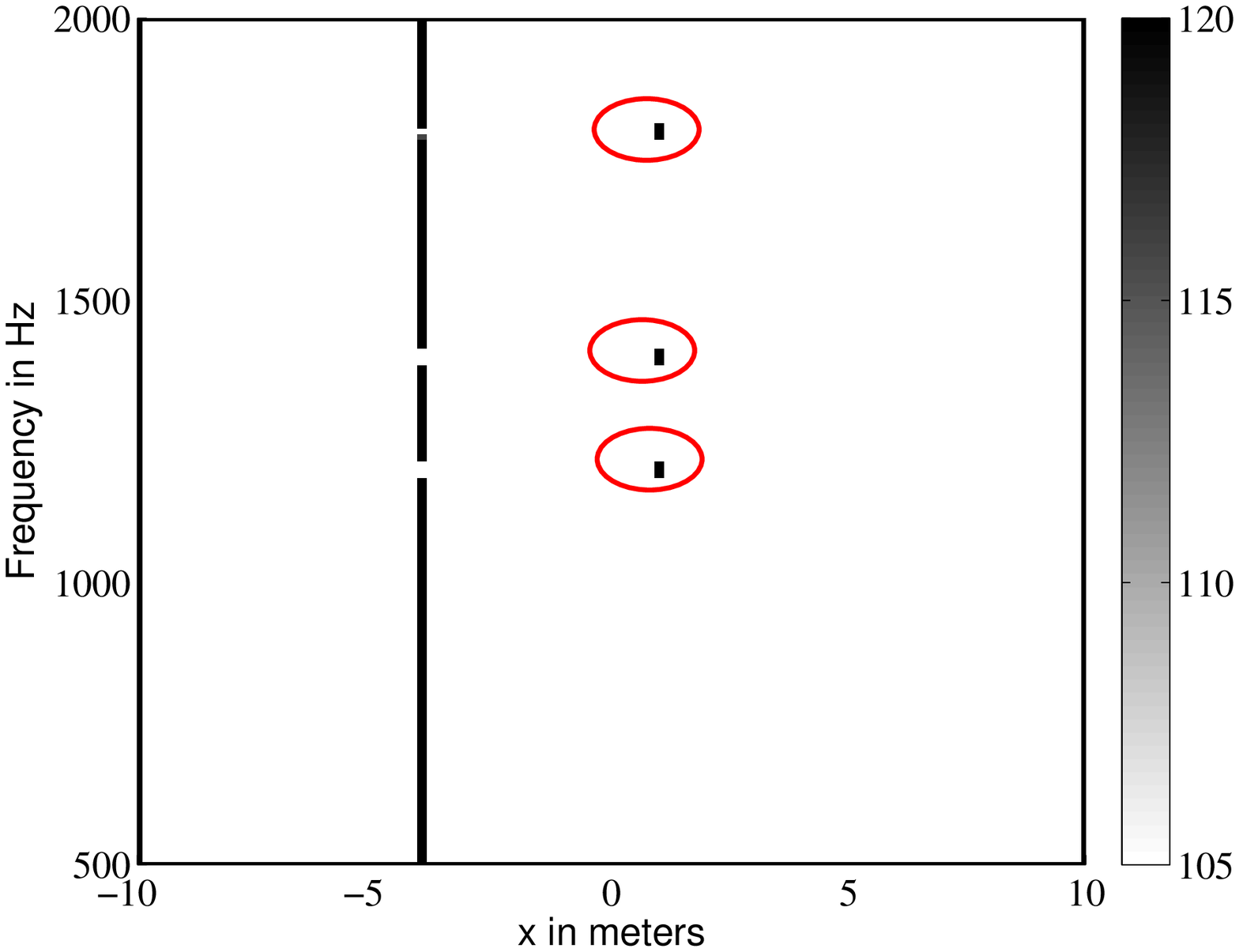}}
\caption{Localization in the frequency-distance domain obtained (in the case without noise). (a) Initial BF-MS. (b) DAMAS-MS. (c) SDM.  (d) NR-SOOT (the dynamic ranges shown are 15 dB).\label{random_cosinus_-14_-9_2D_nonnoise}\vspace*{-2mm}}
\end{figure}
\begin{figure}[h!]
\centering
\subfigure[Initial BF-MS]{\includegraphics[width=7.5cm, height=5cm]{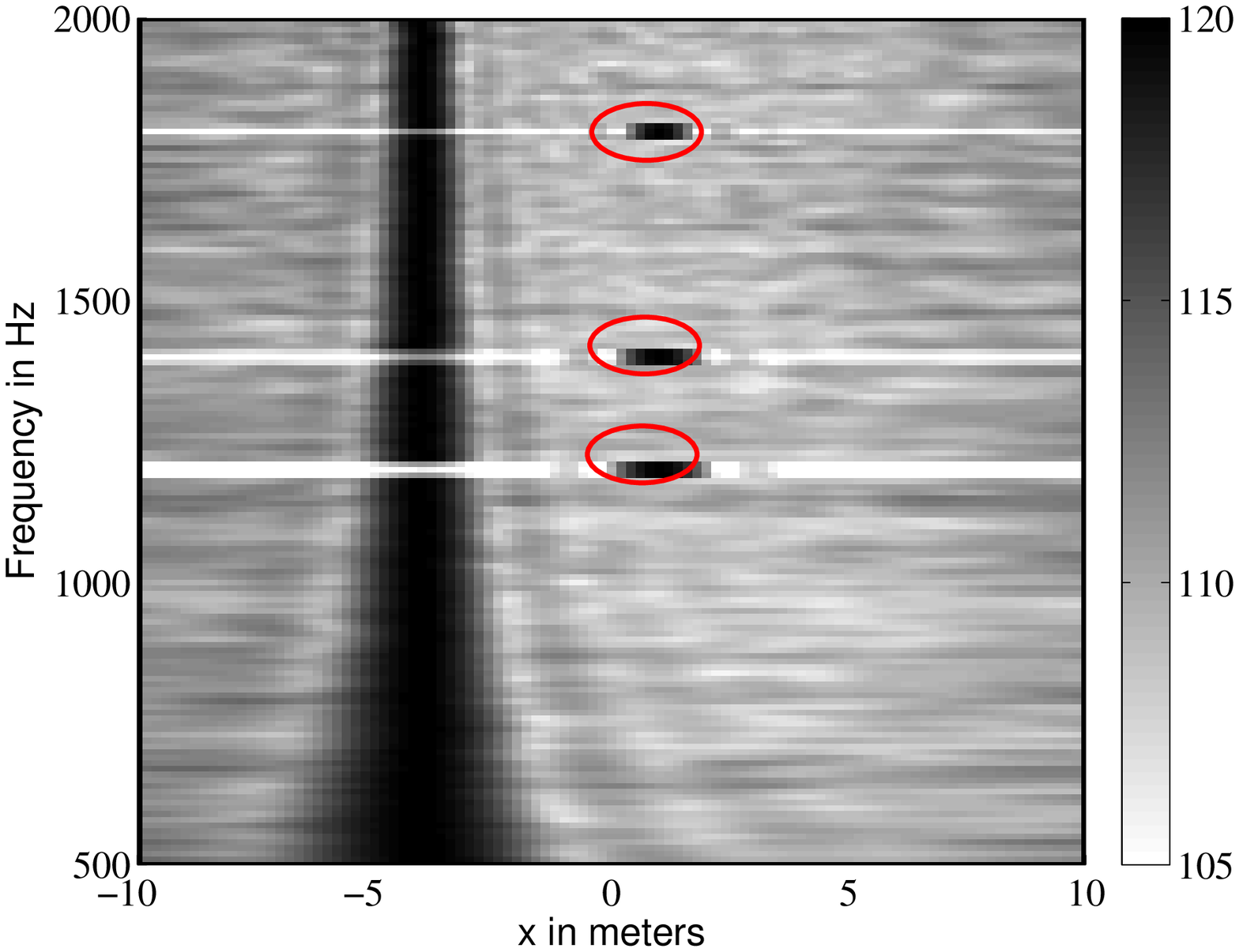}}
\subfigure[Deconvolution with DAMAS-MS]{\includegraphics[width=7.5cm, height=5cm]{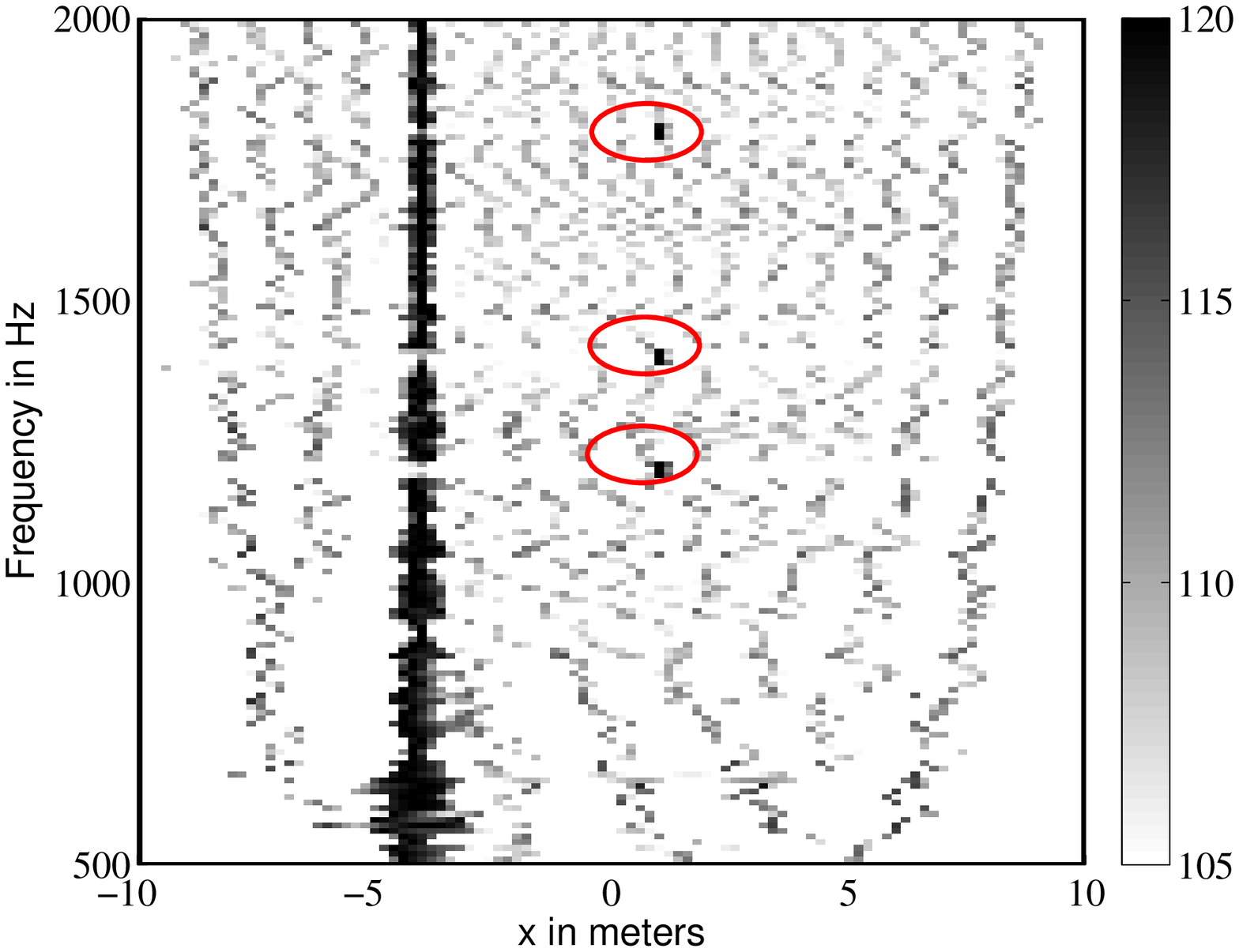}}
\subfigure[Deconvolution with SDM]{\includegraphics[width=7.5cm, height=5cm]{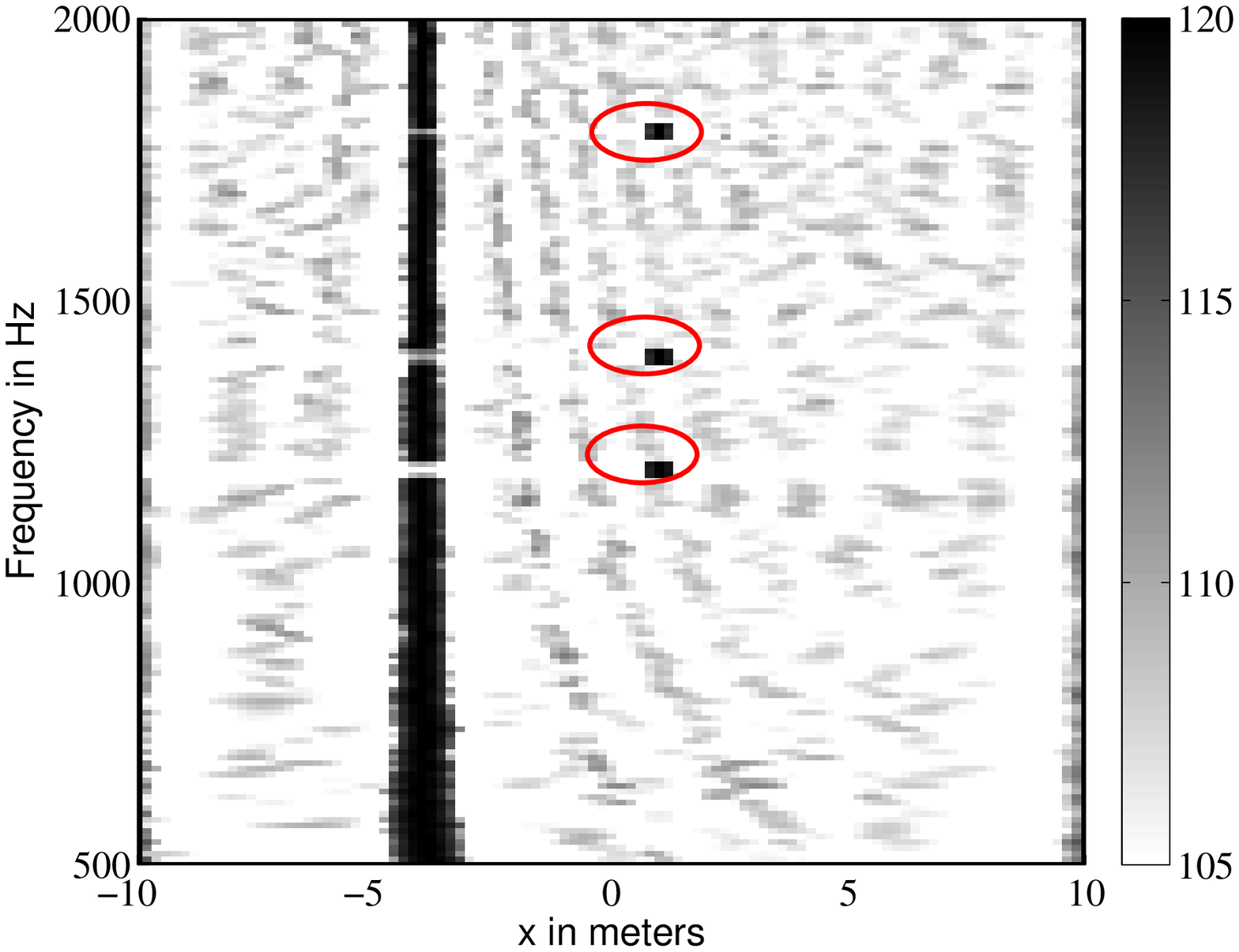}}
\subfigure[Blind deconvolution with NR-SOOT]{\includegraphics[width=7.5cm, height=5cm]{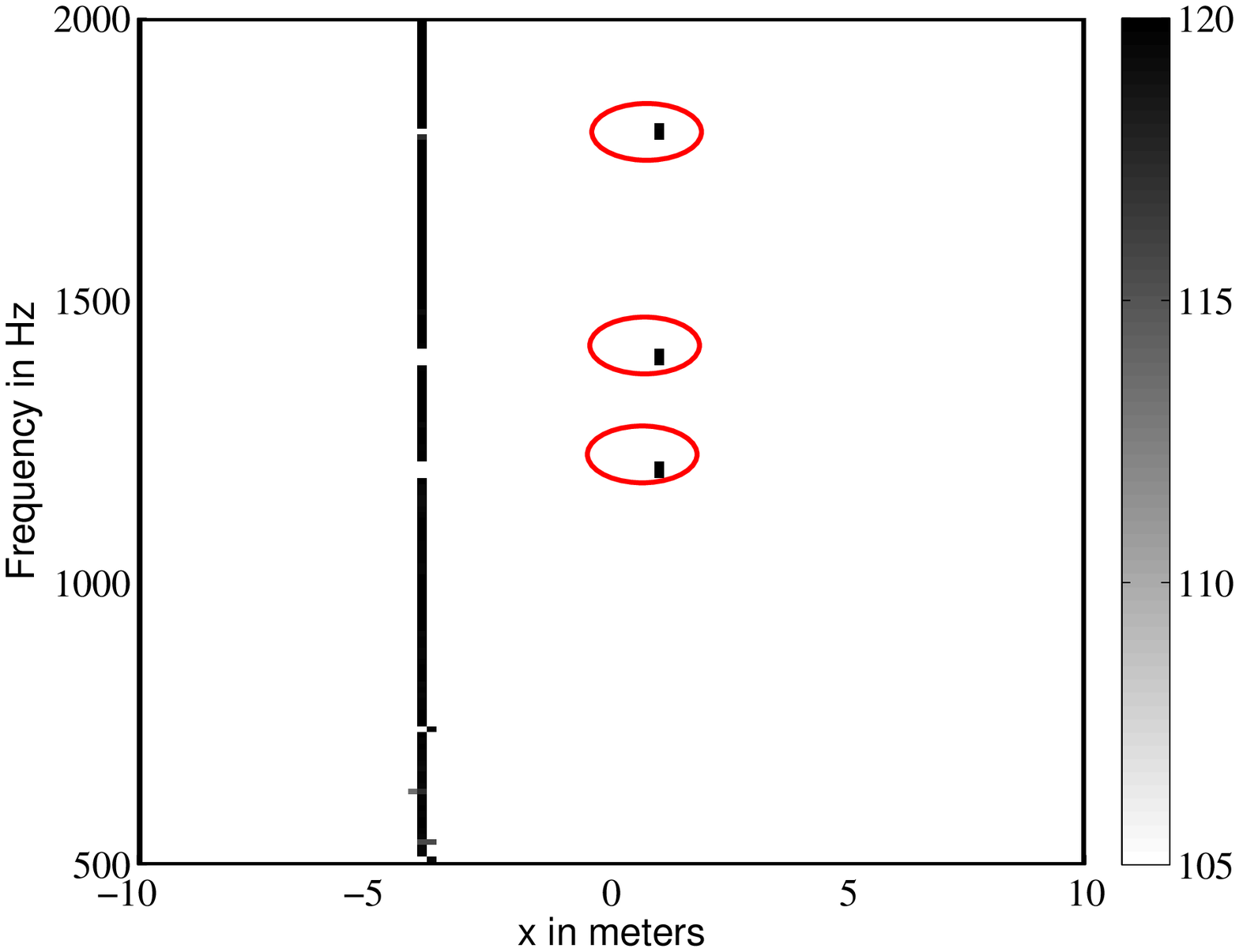}}
\caption{Localization in the frequency-distance domain obtained (with SNR of -5 dB). (a) Initial BF-MS. (b) DAMAS-MS. (c) SDM. (d) NR-SOOT (the dynamic ranges shown are 15 dB).\label{random_cosinus_-14_-9_2D_noise}}
\end{figure}

The two-dimensional localization maps \bn{are shown in Figure~\ref{random_cosinus_-14_-9_2D_nonnoise}} (without noise) and Figure~\ref{random_cosinus_-14_-9_2D_noise} (SNR of -5 dB), \bn{with each Figure showing the initial BF-MS and the results obtained by DAMAS-MS, SDM, and NR-SOOT}.
\bn{For the case without noise of Figure~\ref{random_cosinus_-14_-9_2D_nonnoise}, all of the methods improve the BF-MS output, localize the two sources, and allow identification as one broadboand source and a sum-of-sine source. 
However, the results obtained using DAMAS-MS and SDM are not as good as those using the NR-SOOT algorithm, because the source localizations are spread over several $x$ positions.} 
Moreover, by studying the different zones indicated in the red ellipses in Figure~\ref{random_cosinus_-14_-9_2D_nonnoise} and Figure~\ref{random_cosinus_-14_-9_2D_noise}, which are related to the autospectrum of the sine source at the three frequencies of $1200\, Hz$, $1400\, Hz$, and $1800\, Hz$, some other conclusions can be drawn.
 The results obtained using the DAMAS-MS method are not performing, as some noise appears, as indicated by the red arrows (Fig.~\ref{random_cosinus_-14_-9_2D_nonnoise}b).
For the case of a SNR of -5 dB (Fig.~\ref{random_cosinus_-14_-9_2D_noise}), 
DAMAS-MS and SDM do not identify the sources and give several false alarms. 
On the contrary, the NR-SOOT algorithm still gives good results and provides the best performance compared to the two other methods.    

Consequently, the NR-SOOT algorithm has better performances than the DAMAS-MS and SDM methods in terms of localization, 
as the source $S_2$ is spread over several $x$ positions by DAMAS-MS and SDM, whereas the proposed NR-SOOT algorithm manages to estimate a point source at the true source position. 

\subsection{Real data}
\begin{figure}[h!]
\centering
\subfigure[Initial BF-MS]{\includegraphics[width=7.5cm, height=5cm]{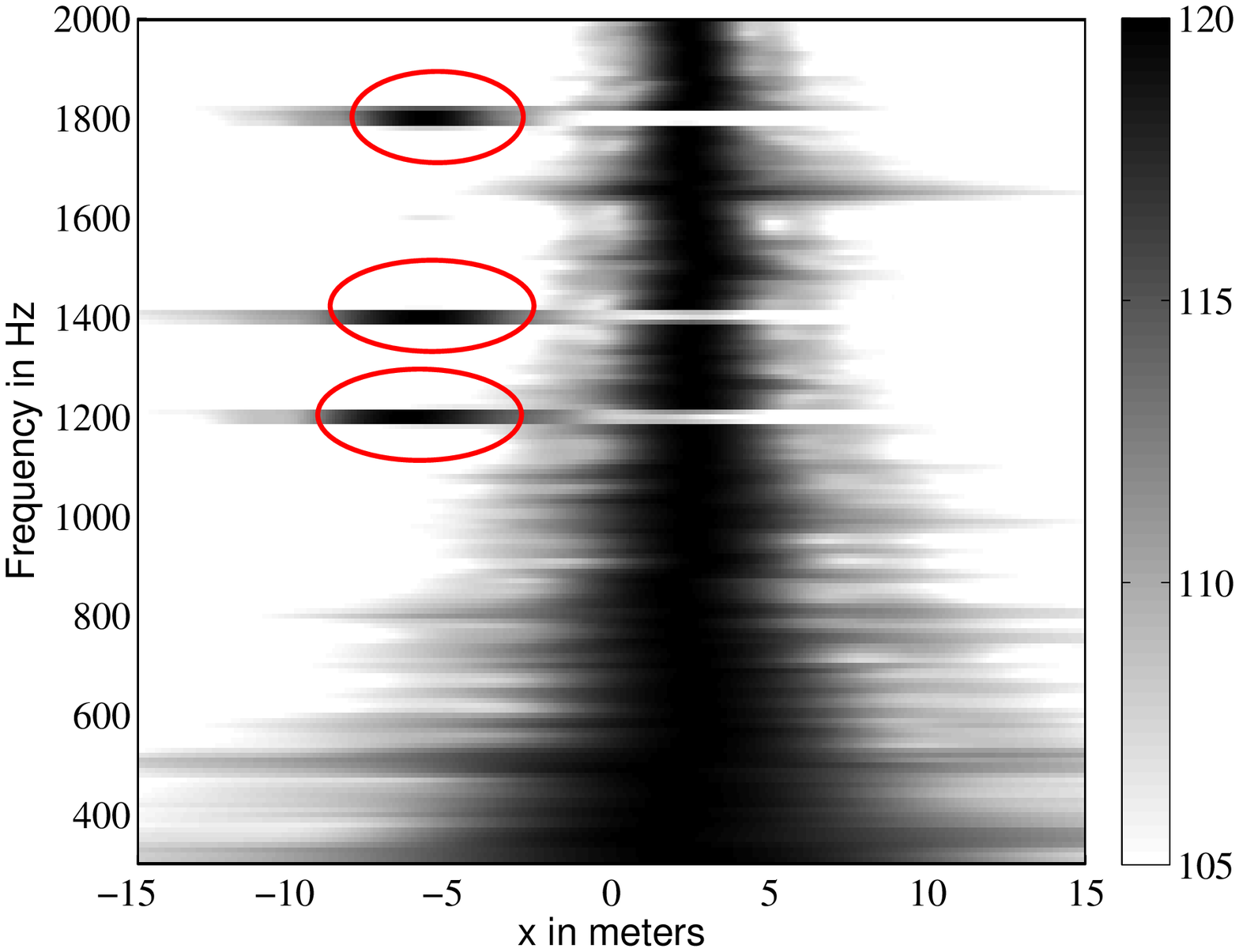}}
\subfigure[Deconvolution with DAMAS-MS]{\includegraphics[width=7.5cm, height=5cm]{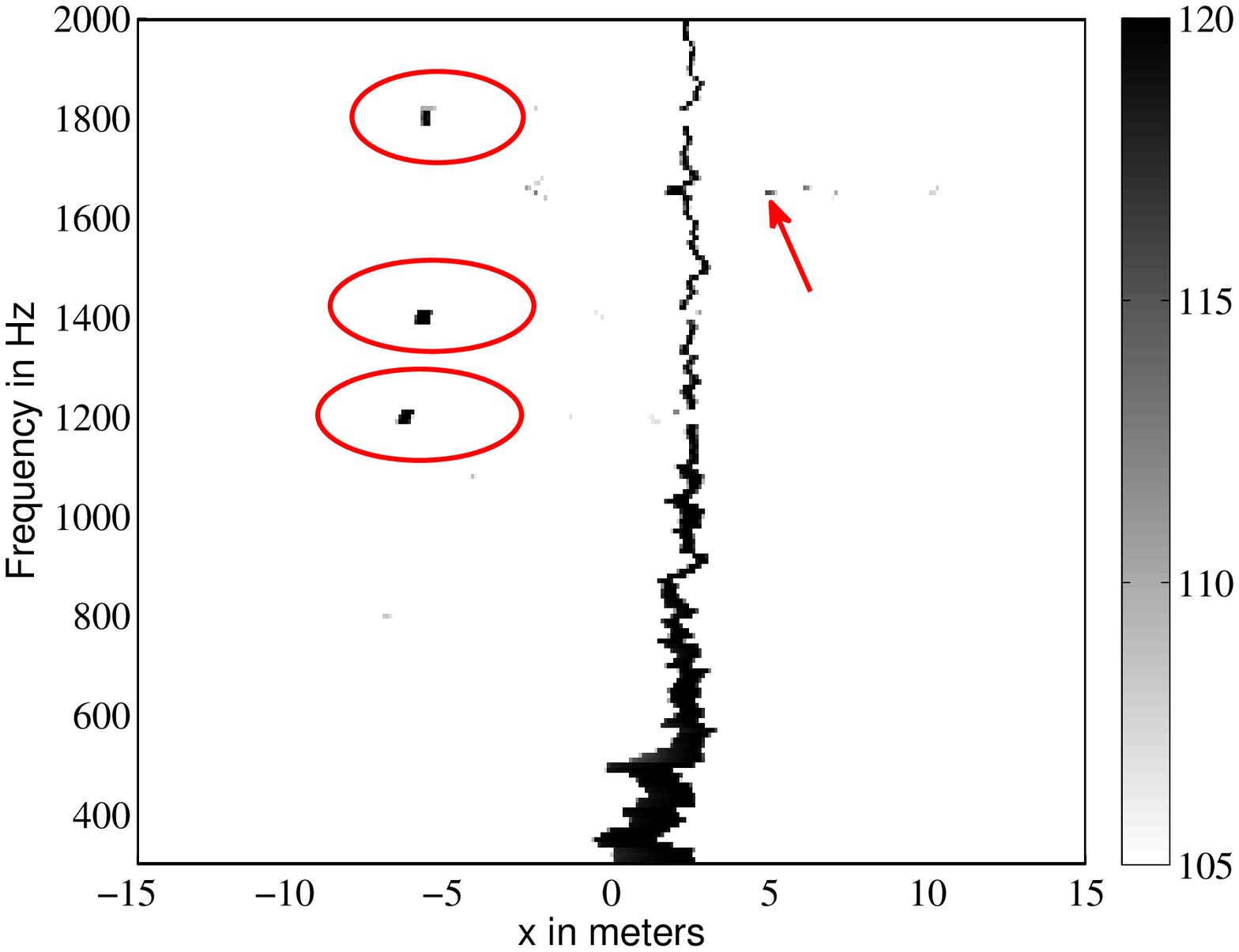}}
\subfigure[Deconvolution with SDM]{\includegraphics[width=7.5cm, height=5cm]{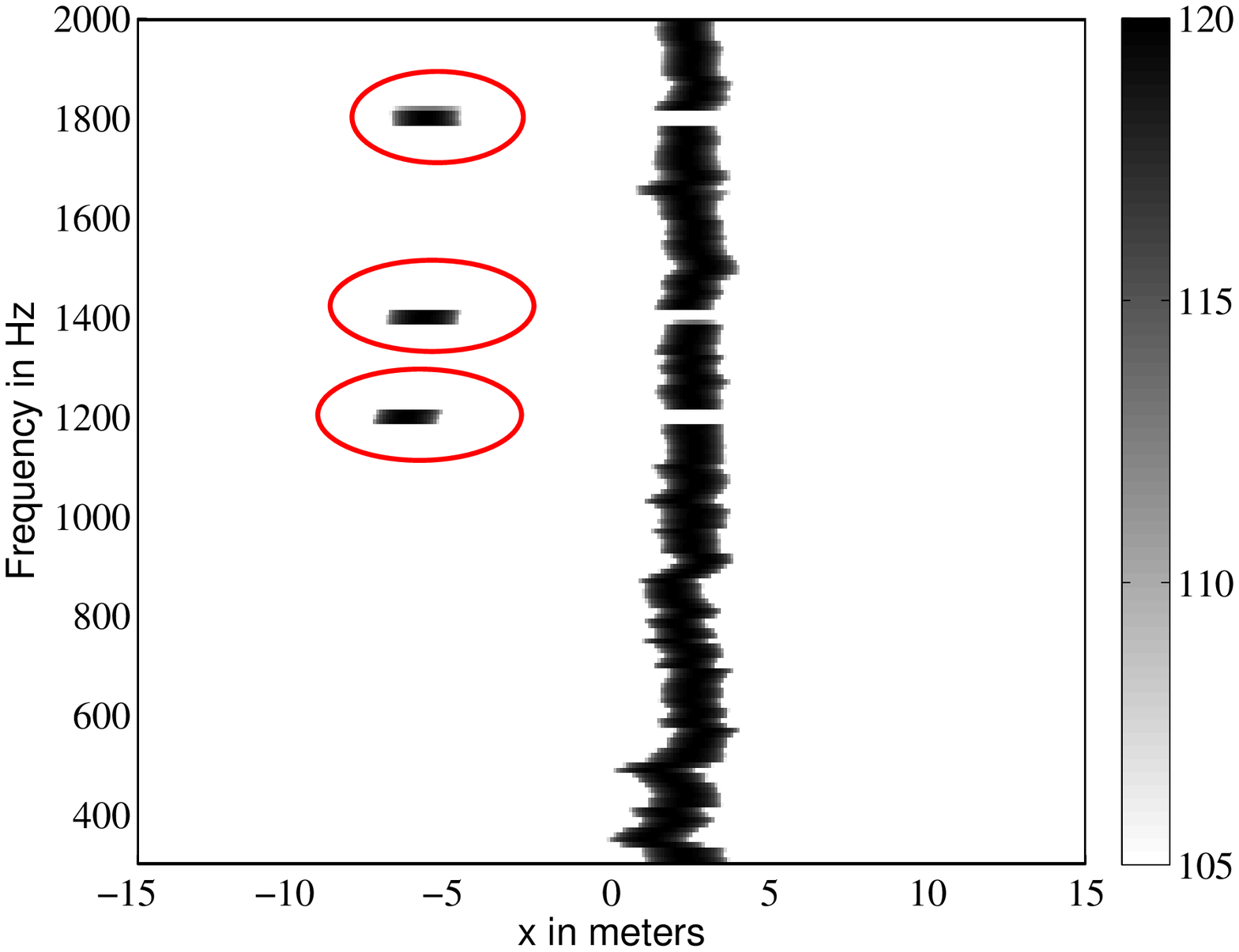}}
\subfigure[Blind deconvolution with NR-SOOT]{\includegraphics[width=7.5cm, height=5cm]{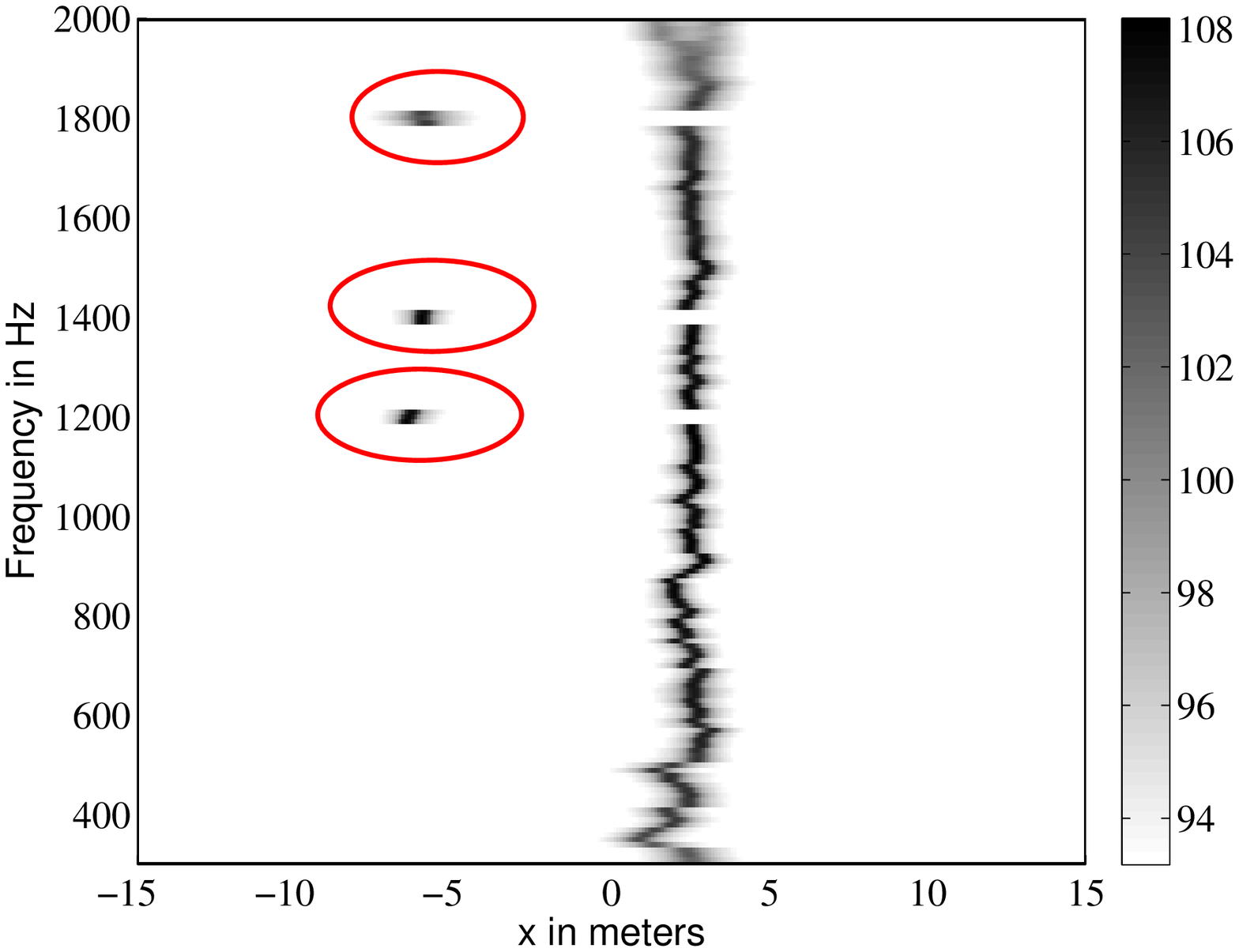}}
\caption{Localization obtained in the frequency-distance domain for the model ship with two artificial sources, traveling at 2 m/s. (a) Initial BF-MS. (b) DAMAS-MS. (c) SDM. (d) NR-SOOT (the dynamic ranges shown are 15 dB).\label{config10_1_191ms}}
\end{figure}

\begin{figure}[h!]
\centering
\subfigure[Initial BF-MS]{\includegraphics[width=7.5cm, height=5cm]{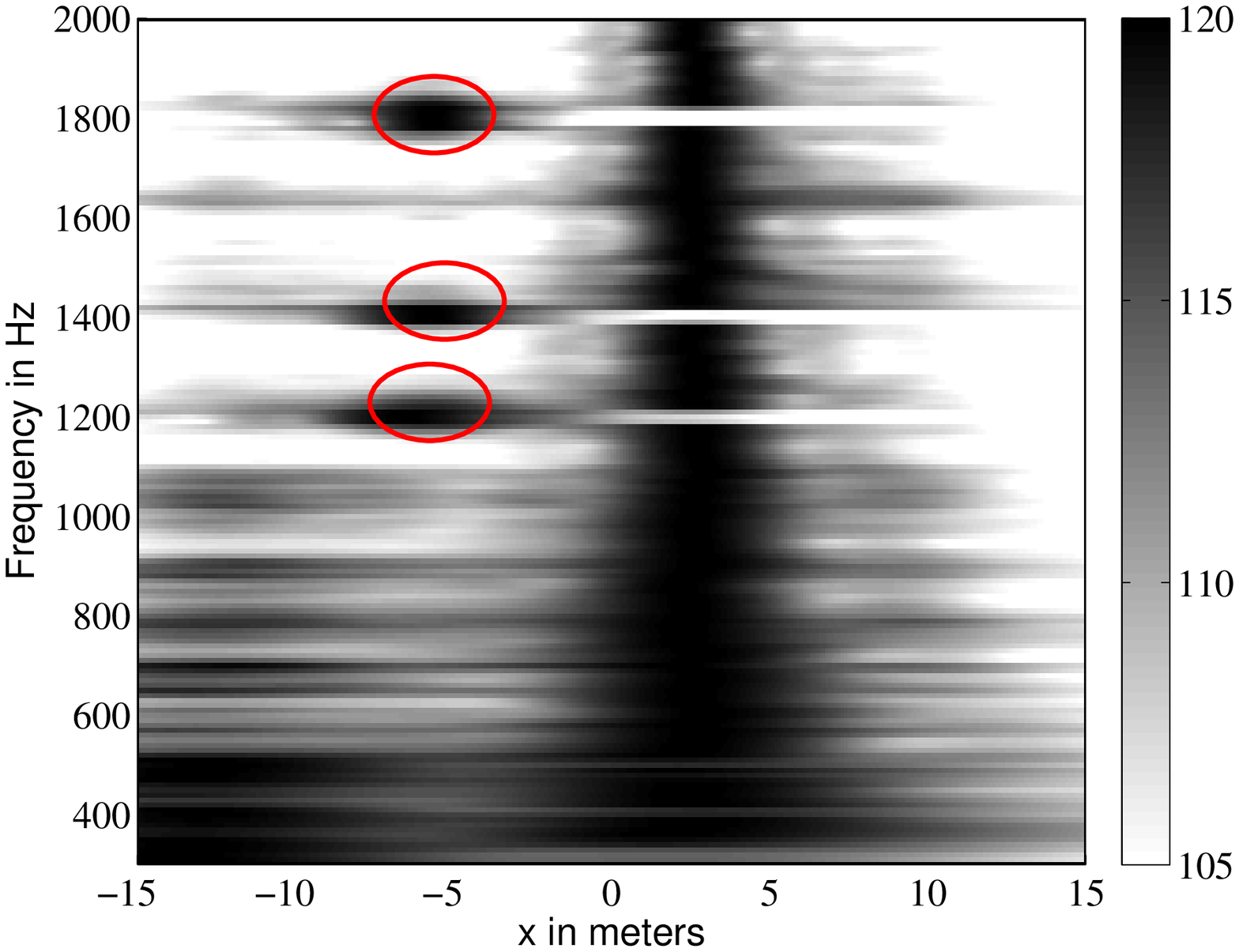}}
\subfigure[Deconvolution with DAMAS-MS]{\includegraphics[width=7.5cm, height=5cm]{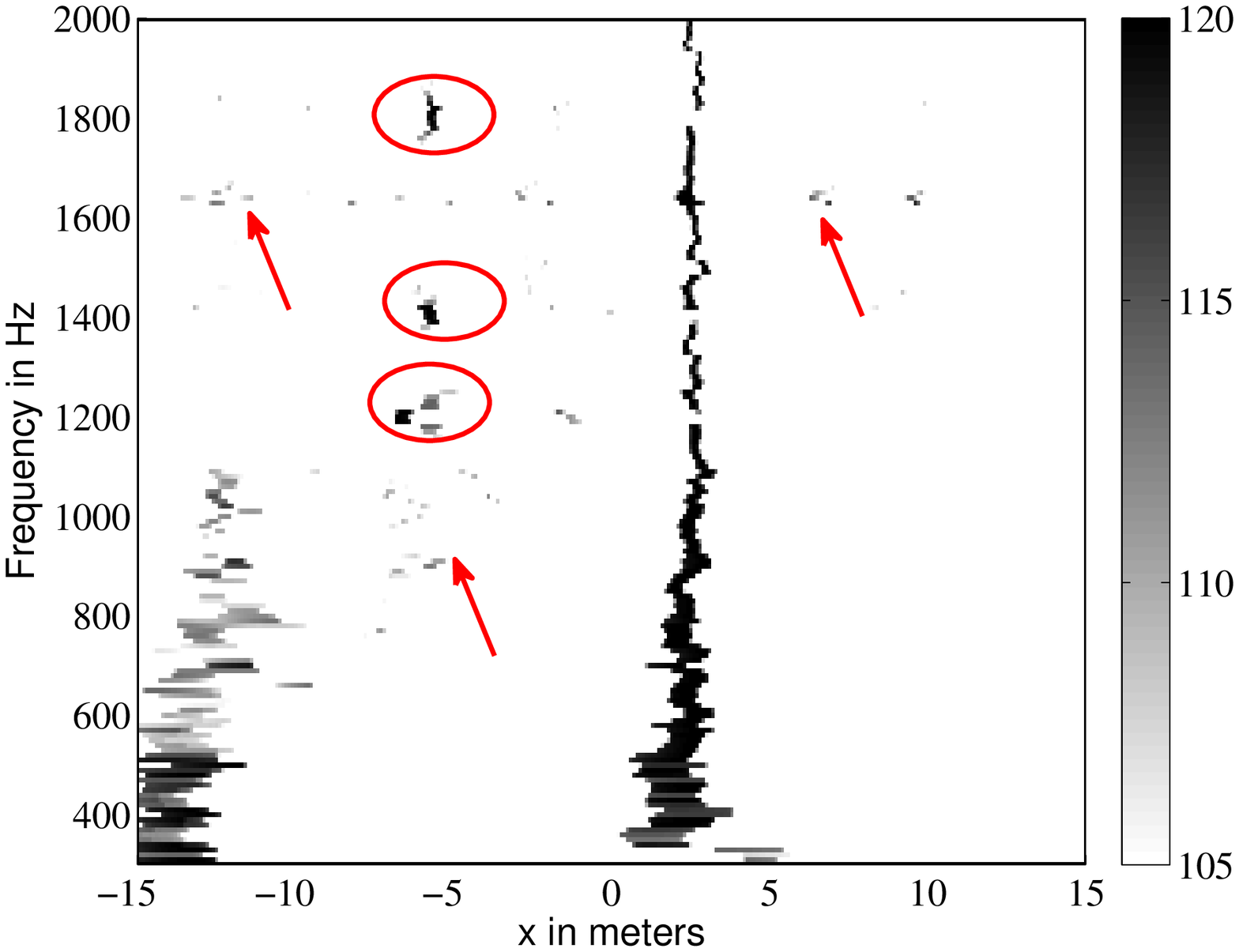}}
\subfigure[Deconvolution with SDM]{\includegraphics[width=7.5cm, height=5cm]{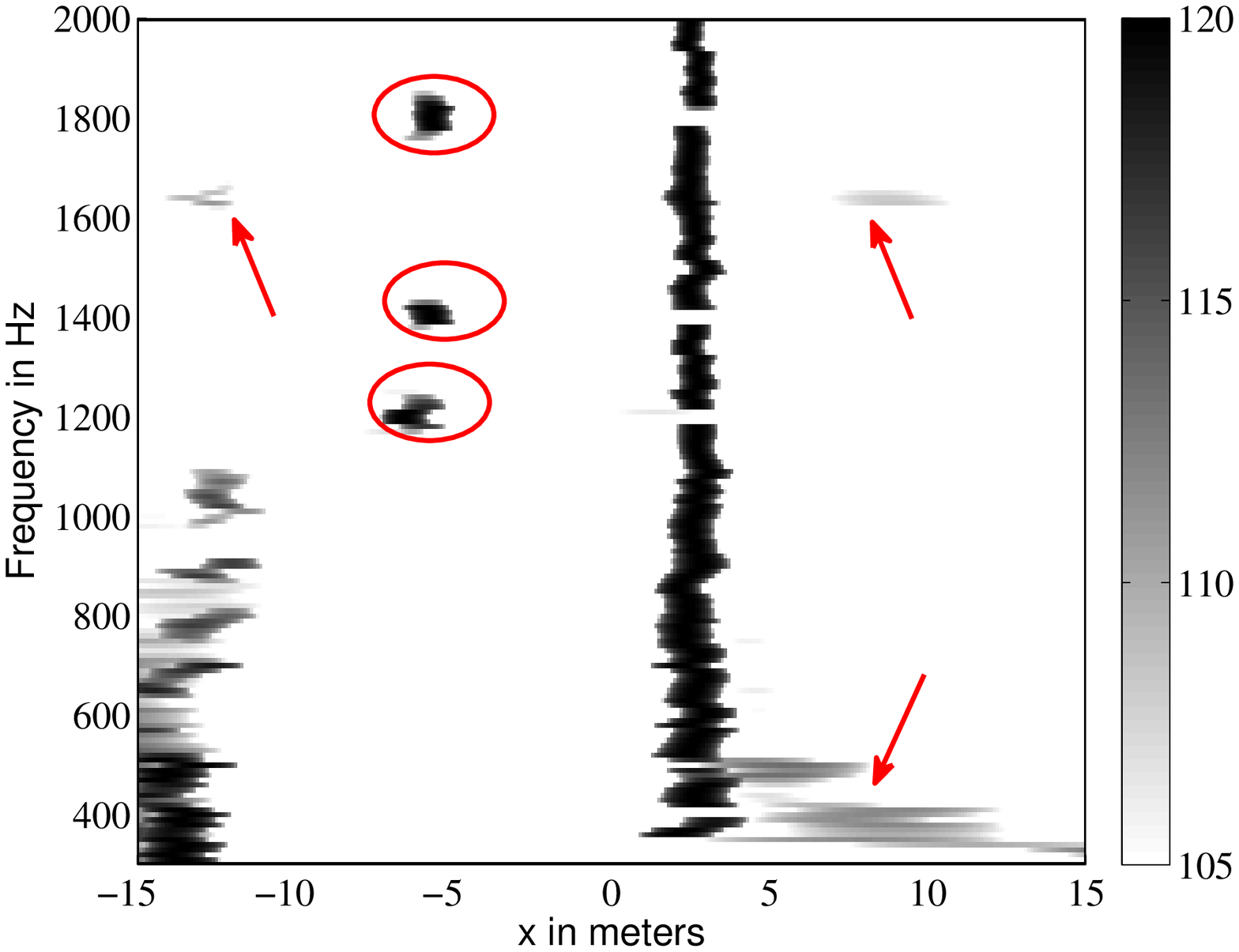}}
\subfigure[Blind deconvolution with NR-SOOT]{\includegraphics[width=7.5cm, height=5cm]{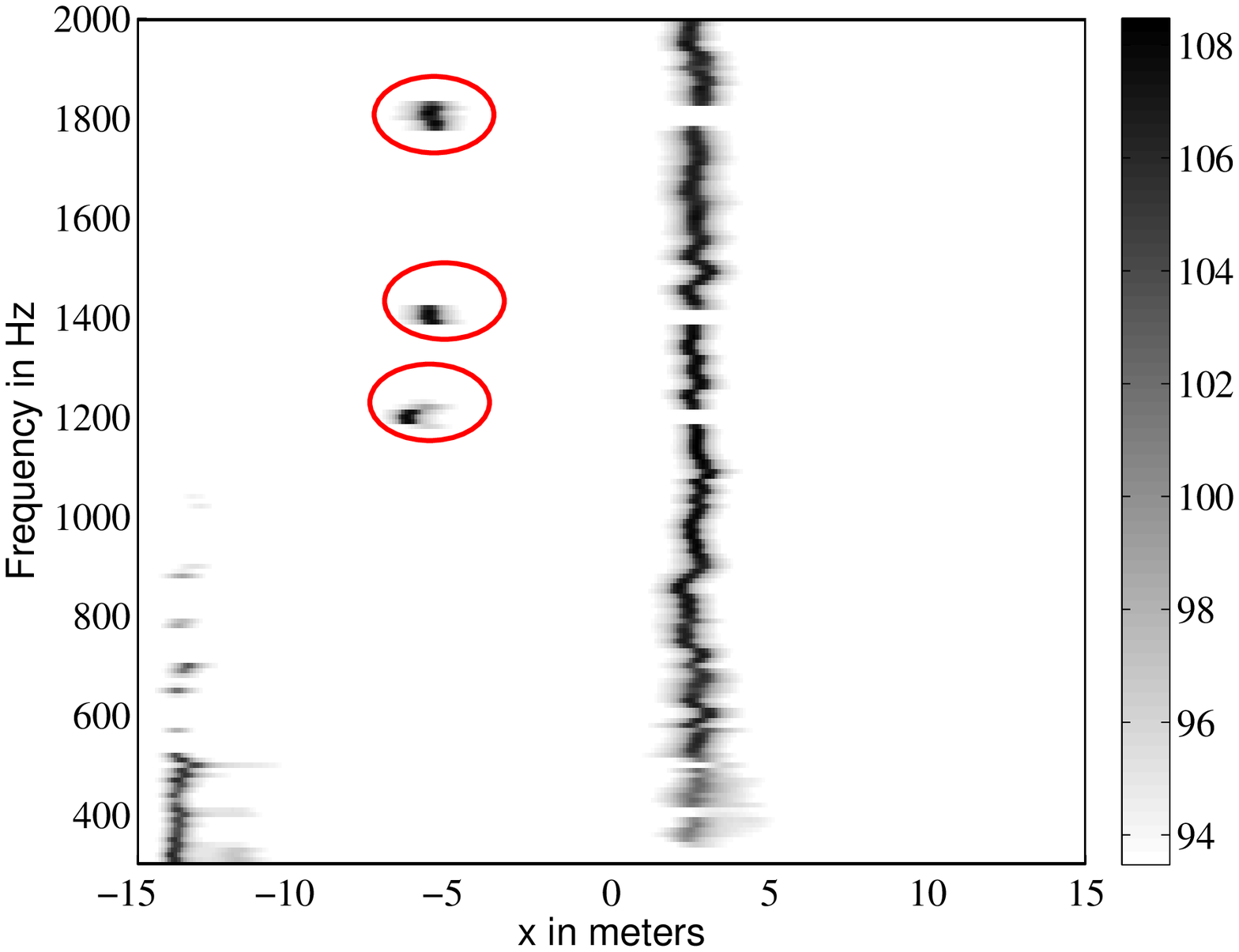}}
\caption{Localization obtained in the frequency-distance domain for the model ship with two artificial sources, traveling at 5 m/s. (a) Initial BF-MS. (b) DAMAS-MS. (c) SDM. (d) NR-SOOT (the dynamic ranges shown are 15 dB).\label{config10_3_520ms}}
\end{figure}

We finally compare the proposed NR-SOOT algorithm with the classical methods using real data. 
The experiment was conducted in January 2015 by DGA naval systems at Lake Castillon, a mountain lake in the French Alps with an average depth of $100\,m$ and a maximum width of $600\,m$. 
This consisted of towing a 21-m-long scale model of a surface ship. The ship
hull included two shakers, $S_1$ and $S_2$, that generated two point acoustic sources outside the hull: a sum of 3 sine functions at frequencies of $1200\, Hz$, $1400\, Hz$, and $1800\, Hz$, located at $x = -5.9 m$, and
a random broadband source located at $x = 2.3 m$. A linear antenna of nine hydrophones that were equally spaced by $0.5\, m$ recorded the propagated acoustic signals over $D = 14.15\, s$ for the source speed of $v = 2\,m/s$ (Fig.~\ref{config10_1_191ms}). 
We also consider the same configuration with the source speed of $v = 5\,m/s$ over $D = 5.3\, s.$ (Fig.~\ref{config10_3_520ms}). The coordinate system of the
array was used to describe all of the geometries, with the origin corresponding to the array center. The array was immersed at $10\,m$ in depth and was positioned
at $2.50\,m$ from the closest point of approach in the $y$ direction. The source trajectory is calculated using a tachymeter system on the idler pulley.
The acquisition time considered for the array processing is sufficient, such that the ship model passed by entirely above the antenna. 
In these Figures, the zones indicated in the red ellipses correspond to the estimated autospectrum of the sine source at the three frequencies of $1200\, Hz$, $1400\, Hz$, and $1800\, Hz$, and the red arrows show the remaining noise or the false alarms.\\
First, we consider the results in the case of the source speed $v =  2\,m/s$ (Fig.~\ref{config10_1_191ms}). 
The three methods improve the BF-MS output and identify the sources. DAMAS-MS and NR-SOOT have better performances than SDM in terms of localization. 
However, for the result of the DAMAS-MS method, there are some false alarms that are indicated by the red arrows in Figure~\ref{config10_1_191ms}b.\\  
\bn{Secondly, we consider the case with the source speed $v = 5\,m/s$, for which the signal in the recording is more noisy. 
In this configuration, one new `natural' source 
appears at the wake of the ship (Fig.~\ref{config10_1_191ms}d, bottom left). 
Three methods identify three sources, whereby the sine source is better localized by the NR-SOOT algorithm than the other methods. 
Both the DAMAS-MS and SDM methods show many false alarms, which are indicated by the red arrows in Figure~\ref{config10_3_520ms}b, c. In particular, the localization of the `natural' source is only possible with the NR-SOOT algorithm.}
In conclusion, our results from this experiment remain true to our hypothesis, as well as our predictions. 
The results shown in Figure~\ref{config10_1_191ms} and Figure~\ref{config10_3_520ms} present the best results with perfect source location and improved  
robustness against noise for the NR-SOOT algorithm.

\section{Conclusions}
This paper \bn{proposes a new method, known as NR-SOOT, that is an extension} of the SOOT algorithm \cite{Reppetti_A_2014_j-ieee-spl_sparse_bdsl1l2r}, for moving-source 
localization \bn{based on blind deconvolution} in underwater acoustic data. As the number of sources is small enough \bn{and they do not spread spatially}, its autospectrum has a sparse representation, 
and it is possible to obtain more accurate results for blind deconvolution through a regularization function. The smooth approximation of $\ell_1/\ell_2$ shows very good performances in terms of localization and suppression of false alarms, \bn{and provides better results than DAMAS-MS and SDM, particularly for low SNRs.}  

\section*{Acknowledgements}
\label{sec:acknowledgements}
This work was financially supported by the Minist\`ere du Redressement Productif (Direction G\'en\'erale de la Comp\'etitivit\'e, de l'Industrie et des Services) and by the DGA-MRIS, grant RAPID ARMADA N\textsuperscript{o}122906030.

\end{document}